\documentclass[a4paper,11pt]{article}
\pdfoutput=1 

\usepackage{jheppub}
\usepackage{epsfig}
\usepackage{amsmath}
\usepackage{graphicx}
\usepackage{amssymb,mathtools,booktabs,array,tabularx,microtype,float,needspace}

\def\be{\begin{equation}}
\def\ee{\end{equation}}
\def\bea{\begin{array}}
\def\eea{\end{array}}
\def\beqa{\begin{eqnarray}}
\def\eeqa{\end{eqnarray}}
\def\beqas{\begin{eqnarray*}}
\def\eeqas{\end{eqnarray*}}

\def\bp{\begin{picture}}
\def\ep{\end{picture}}
\def\bc{\begin{center}}
\def\ec{\end{center}}
\def\bfig{\begin{figure}}
\def\efig{\end{figure}}

\def\bit{\begin{itemize}}
\def\eit{\end{itemize}}
\def\nn{\nonumber}
\def\f{\frac}

\def\[{\left[}
\def\]{\right]}
\def\({\left(}
\def\){\right)}

\def\..{\left.}
\def\.{\right.}

\def\ra{\rightarrow}
\def\la{\leftarrow}

\def\tm{\times}

\def\ev{\rm eV}

\def\da{\dagger}

\def\la{\lambda}

\def\ep{\epsilon}

\newcommand{\SU}{\mathrm{SU}}
\newcommand{\SO}{\mathrm{SO}}

\newcolumntype{Y}{>{\raggedright\arraybackslash}X}
\newcolumntype{L}[1]{>{\raggedright\arraybackslash}p{#1}}
\newcommand{\GeV}{\,\mathrm{GeV}}
\newcommand{\TeV}{\,\mathrm{TeV}}

\newcommand{\keV}{\,\mathrm{keV}}
\newcommand{\MeV}{\,\mathrm{MeV}}

\hypersetup{colorlinks=false,pdfborder={0 0 0},pdftitle={Higgsino dark matter explanation of LZ high-recoil event from deflected anomaly mediation with $U(1)_X$ generated $\mu$ term},pdfauthor={}}
\title{Higgsino dark matter explanation of LZ high-recoil event from deflected AMSB with $U(1)_X$ generated $\mu$ term}
\author{Ke-Lin Li, Fei Wang}
\affiliation[a]{School of Physics, Zhengzhou University, Zhengzhou 450000, P. R. China}
\emailAdd{feiwang@zzu.edu.cn}
\abstract{ Motivated by the recent LZ high-recoil event, we propose to study the Higgsino dark matter in a supersymmetric $U(1)_X$ framework with one charged visible singlet. The gauge symmetry
forbids the bare Higgs bilinear and the cubic $S^3$ while allowing the
singlet interaction that generates an effective Higgsino mass after
symmetry breaking.
We present the six-state neutralino mixing matrix, including the
$U(1)_X$ gaugino and the singlino, and the resulting neutral Higgsino
mass splitting.
Different deflected-anomaly-mediated parameter regimes provide the
soft-mass hierarchies needed for light, 1.1 TeV and heavy Higgsino
candidates. 
The cosmological
consequences of the messenger sector in deflected AMSB can affect greatly the thermal history of the dark matter. 
Late messenger decays can supplement an underabundant
thermal population or dilute an overabundant one through entropy
production. These possibilities are governed by a common final-yield
relation: a dominating messenger must have sufficiently suppressed
Higgsino production for net dilution to succeed. We organise the
500 GeV, 1.1 TeV and 100 TeV reference regimes around this condition.
Direct-detection and solar-capture constraints are discussed as tests
of the resulting spectrum and final abundance. The elastic bino- or singlino-like interpretation of the same event is also 
realized in this spectrum: the Higgsino asymmetry $|X_A|\simeq10^{-2}$ that a $248\keV$ recoil requires arises without tuning, 
with a spin-dependent rate lying close to the present experimental sensitivity.}
\begin{document}
\maketitle
\clearpage

\section{Introduction}
\label{sec:intro}
The LUX-ZEPLIN (LZ) extended-recoil analysis reports a candidate at
$E_R=248\pm23\,\mathrm{(stat)}\pm23\,\mathrm{(sys)}\keV$ in an exposure
of $2.84$ tonne-year, with global and local significances of $2.6\sigma$
and $3.4\sigma$, respectively~\cite{LZext}. Although the collaboration does 
not identify the event as dark matter with this significances, the unusually 
large recoil energy makes the event a useful probe of dark-matter scenarios, 
for example, the Higgsino dark matter scenario.

A nearly pure Higgsino provides a simple connection between electroweak
symmetry breaking and dark matter. Its gauge interactions determine the
leading annihilation channels and the off-diagonal neutral-current
interaction of its two neutral Majorana components. Thermal freeze-out with full dark matter relic density
requires its mass close to $1.1\TeV$, whereas non-thermal histories allow
lighter or heavier Higgsinos. A small neutral Higgsino mass splitting can make
nuclear scattering inelastic and connect the neutralino spectrum to
high-recoil direct-detection searches~\cite{FanReece,Freese,Graham}.

Three issues must be addressed together. First, the neutral Higgsino
splitting depends on the gaugino masses and on additional neutral-fermion
mixing. Away from cancellations, the estimate
$M_{\rm gaugino}\sim m_Z^2/\Delta m_\chi$ places the electroweak gauginos
well above the Higgsino mass for a sub-MeV splitting~\cite{FanReece,DuWang}.
Second, the inelastic interaction responsible for a terrestrial recoil
also enters solar capture. Under the thermalization and halo assumptions, the splitting favored by the thermal-Higgsino interpretation of the event is in tension with solar-neutrino searches~\cite{Pospelov,Bose}. In fact, a dedicated likelihood analysis that includes IceCube limits on solar annihilation finds compatibility over masses from about $2\TeV$ to several hundred $\TeV$ with splittings of roughly $350$--$525\keV$, whereas the $1.1\TeV$ thermal point is only marginally allowed and dependent on the assumed high-speed tail, the solar cooling channels and the treatment of the high-energy
sideband~\cite{Freese:2026b}. Third, a non-thermal history must account for both the entropy released by
a late source and the dark matter produced in its decay.

 Besides the attractive inelastic Higgsino scenario, the event has a large number of other interpretations, which include elastic
bino- or singlino-like neutralinos (whose diagonal axial $Z$ coupling is set by
the Higgsino asymmetry rather than by the total Higgsino fraction
~\cite{Roy:2026}), high-scale supersymmetry
and Peccei--Quinn dynamics~\cite{Yin,Langhoff}, bino--wino cancellations~\cite{DuWang},
singlino mixing~\cite{Bisal}, modified solar thermalisation~\cite{Wu},
or non-thermal dark-matter production~\cite{Rodd}, freeze-in production of inelastic dark matter~\cite{FreezeIn}, 
relativistic boosted dark matter~\cite{BoostedCR}, dark-QCD
states~\cite{DarkQCD} etc. The two kinematically distinct classes, endothermic inelastic
scattering and relativistic boosted scattering, have been shown to be spectrally
separable across the high-energy window. The inelastic spectra place only
$6$-$17\%$ of events below $150\keV$~\cite{Mahapatra:2026} so that a few
additional events would distinguish them.

 Our focus is the $U(1)_X$ approach to the SUSY $\mu$ problem, studying the corresponding neutralino mass mixing 
 and its possible realization of small Higgsino splitting.  Besides, we discuss its realization from deflected AMSB, 
 showing the connection between different deflected-AMSB~\cite{Pomarol:1999ie,Fei:new} spectra and late messenger decays.
 Possible cosmological consequences of a messenger sector in deflected-AMSB, which received little attention 
and can affect dark matter relic abundance, is another focus of this work. 
The elastic bino- or singlino-like interpretation of the event can also be realized within this framework. 
The extra $U(1)_X$ $D$-term and the singlino admixtures push the diagonal axial $Z$ coupling, 
i.e.\ the Higgsino asymmetry $X_A=|N_{13}|^2-|N_{14}|^2$, to the value $|X_A|\sim10^{-2}$ that the $248\keV$ recoil requires, 
without tuning. The resulting spin-dependent rate then lies close to the present experimental sensitivity.
Note that we use the LZ event as a motivation rather than as evidence for any particular model.

We adopt a gauged $U(1)_X$ extension in which $H_u$, $H_d$ and a single
visible singlet $S$ all carry $U(1)_X$ charge. The gauge symmetry itself
forbids the bare bilinear $\mu H_uH_d$ and the cubic $S^3$, while allowing the 
$\lambda SH_uH_d$ term. The singlet vacuum then breaks the extra symmetry and
generates $\mu_{\rm eff}=\lambda\langle S\rangle$. The $\mu$ problem is
therefore solved by the Abelian gauge structure alone, with no additional
discrete symmetry and no second singlet. Because $S$ is charged and has no
superpotential cubic, the singlet mass is not an independent parameter: it
is tied to the Abelian breaking scale, and the active neutralino sector
closes at six states. Without a Green--Schwarz contribution, the $U(1)_X$ charge assignment used here is anomaly-free
 with or without an $E_6$ embedding.

The subsequent analysis has two connected but distinct parts.
Deflected anomaly mediation sets the gaugino and scalar soft terms,
with the deflection parameter changing mass ratios and relative signs.
The singlet parameters and vacuum then determine whether the lightest
 neutralino is Higgsino-like. The cosmological consequences of the messenger sector in 
 deflected AMSB have received little attention so far, yet they play an important role in this work. Messenger lifetimes, 
 abundances and decay
channels govern the cosmological modification of that spectrum's
thermal relic abundance. A late decay can both produce Higgsinos and
increase the entropy. Its net effect cannot be inferred from the
lifetime alone.

The $500\GeV$, $1.1\TeV$ and $100\TeV$ reference regimes
represent an under-abundant, a full abundant, and an over-abundant thermal dark matter, respectively.
They require different soft-mass and singlet inputs, and need not share
one messenger decay history. We use them to identify the requirements
for non-thermal supplementation or dilution, not necessarily three fully
validated spectra. A sub-MeV splitting and a Higgsino LSP must be
checked simultaneously in the complete neutralino sector.

The paper is organized as follows. Section~\ref{sec:model} introduces
the gauged origin of $\mu$ in the $U(1)_X$ framework and presents the
neutralino matrix and Higgsino splitting. Section~\ref{sec:damsb}
discusses the deflected-AMSB realizations and the relevant messenger-sector.
Section~\ref{sec:relic} discuss the messenger production and entropy dilution
within a common final-yield description. Section~\ref{sec:dd} discusses
direct detection and solar capture, and section~\ref{sec:discussion}
 presents the conclusions. Appendix~\ref{app:soft} gives the deflected-AMSB soft parameters, the
 analytic neutralino reduction.
 Appendix~\ref{app:dilution} provide the related cosmologically details and an
 alternative scalar entropy source.
Experimental reference inputs are in appendix~\ref{app:phenodata}.

\section{A singlet-generated Higgsino mass in the $U(1)_X$ framework}
\label{sec:model}
The Higgsino masses are determined by the $\mu$-parameter, which is a SUSY preserving parameter in the superpotential. However, the $\mu$-parameter is required to be roughly of the order of the soft SUSY breaking scale, as determined by the minimization of the MSSM Higgs potential. Many proposals have been put forward to address the $\mu$-problem, for example, by singlet extension of MSSM to Next-to-Minimal Supersymmetric Standard Model~(NMSSM) or by introducing non-renormalizable terms in the superpotential via Kim-Nilles mechanism~\cite{KN} (or in the Kahler potential via Giudice-Masiero mechanism~\cite{GM}). Most solutions involve two steps~\cite{Bae:2019dgg}: First, the appearance of $\mu$-parameter is prohibited, usually by some discrete or continuous symmetry. Second, the $\mu$ term is regenerated via some mechanism such as symmetry breaking. For detailed realizations of these two steps, one can refer to~\cite{Bae:2019dgg}.


As noted in the introduction, we extend the gauge group of MSSM to $SU(3)_c\times SU(2)_L\times U(1)_Y\times U(1)_X$ in this work and introduce a Standard Model singlet. Both Higgs doublets and the new singlet carry $U(1)_X$ charge. The two conditions that set up the $\mu$ problem are
\begin{equation}
 q_{H_u}+q_{H_d}\ne0,\qquad
 q_S+q_{H_u}+q_{H_d}=0,
\label{eq:higgs-bl-charges}
\end{equation}
so that the $U(1)_X$ charge assignments forbid the bare bilinear $\mu_0H_uH_d$ and $\kappa S^3$ term while permit
\begin{equation}
 W_{\rm vis}\supset\lambda SH_uH_d.
\label{eq:discrete-selection}
\end{equation}

We have several choices of anomaly-free $U(1)_X$ charge assignment:
\bit
\item  GUT uniform $U(1)_X$ charge assignment:

 Requiring the $U(1)_X$ charge assignment being universal on each {complete} $SU(5)$ multiplet, the Standard-Model Yukawa couplings can fix the pattern entirely,
\begin{equation}
 q_X(10)=q_X(\bar5)=q,\qquad
 q_X(H_u)=q_X(H_d)=-2q,\qquad
 q_X(S)=4q ,
 \label{eq:gut-uniform-charges}
\end{equation}
because $QH_uu^c$ and $LH_de^c$ invariance require
$q+q(H_u)+q=0$ and $q+q(H_d)+q=0$. The Higgs bilinear carries charge
$-4q\neq0$, so the bare $\mu$ term is forbidden and
$\mu_{\rm eff}=\lambda\langle S\rangle$ is generated only after
$U(1)_X$ breaking. The pattern
\eqref{eq:gut-uniform-charges} is the $U(1)_\psi$ of $E_6$, under which
$$\mathbf{27}=\mathbf{10}_{+1}\oplus\bar{\mathbf5}_{+1}\oplus\mathbf{1}_{+1}
\oplus\mathbf{5}_{-2}\oplus\bar{\mathbf5}_{-2}\oplus\mathbf{1}_{+4}.$$
We should note that the equality $q(10)=q(\bar5)$ used previously is an extra choice rather than a requirement of grand unification. $E_6$ has rank 6 and contains two chains:
$E_6 \supset \SO(10)\times U(1)_\psi$ and $\SO(10)\supset \SU(5)\times U(1)_\chi$.
Hence there are two Cartan directions that commute with $\SU(5)$, namely
$\psi$ and $\chi$. Any $U(1)$ that commute with $\SU(5)$ is a linear combination
of the two $q_X = a\psi + b\chi$, where only the ratio matters and the overall normalization can be absorbed into $g_X$.
The $\bf{27}$ representation can be decomposed under $(\psi,\chi)$ as
\begin{equation*}
{\bf 27} =\underbrace{\bf 10_{(1,-1)} \oplus \bar{5}_{(1,3)} \oplus 1_{(1,-5)}}_{\text{from}~{\bf 16}~\text{of}~SO(10)}
\oplus \underbrace{\bf 5_{(-2,2)} \oplus \bar{5}_{(-2,-2)} \oplus 1_{(4,0)}}_{\text{from } ~\bf{10},~\bf{1}~\text{of}~SO(10) }.
\end{equation*}
Thus the $U(1)_X$ charge assignment can be 
\beqa
&& q_X(10)=a-b,\quad q_X(\bar5)=a+3b,\quad
 q_X(5_H)=-2a+2b,\quad \nn\\ 
 && q_X(\bar5_H)=-2a-2b,\quad q_X(1_S)=4a .
 \label{eq:e6-family}
\eeqa
Both Yukawa couplings $QH_uu^c$ and $LH_de^c$ are automatically invariant for arbitrary $(a,b)$, because the $\mathbf{27}$ multiplet in $E_6$ satisfies ${\bf 27}\otimes {\bf 27} \otimes {\bf 27}\supseteq {\bf 1}$. The bare $\mu$ term is forbidden whenever $a\neq0$.

Anomaly freedom constrains this $U(1)_X$ charge assignment stringently. Obviously, a vector-like pair of complete $SU(5)$ GUT messenger multiplets give null contributions to all six anomaly coefficients relevant to $U(1)_X$ anomaly, which are $$(A_{33X},A_{22X},A_{YYX},A_{YXX},
 A_{XXX},A_{{\rm grav}X}).$$ 
 So, additional exotic matter multiplets should be introduced to cancel the anomaly and the full set of such exotic matter should be vector-like in its $G_{SM}$ representation.
Previous discussions on the charge assignment of a non-anomalous $U(1)$ that forbid the bare $\mu$-term had been given in~\cite{Cvetic:1997,Hundi:2009,Morrissey:2006,Cohen:2008}.
 
Minimal matter contents contain three $\mathbf{27}$-plets of $E_6$, which include
$$3(\mathbf{10}_1+\bar{\mathbf5}_1+\mathbf{1}_1)
+3(\mathbf{5}_{-2}+\bar{\mathbf5}_{-2})+3\cdot\mathbf{1}_4,$$ for which
all six anomaly coefficients vanish exactly. The same $S$ that generates
$\mu_{\rm eff}$ can give mass to the two extra $Y(\mathbf5_{-2})
+\bar{Y}(\bar{\mathbf5}_{-2})$ pairs. Three generation of right-handed neutrinos can appear in the $\mathbf{16}$ representation, 
carrying non-trivial
charge $q_N=a-5b$. 
However, no singlet of charge $-2q_N$ exists to give their Majorana masses while Dirac mass terms are allowed. 
So, Dirac neutrino mass is generated with tiny Yukawa coupling, while an additional 
singlet (possibly from exotic $\overline{\bf 27}_H$ Higgs) or a projection is required for seesaw-type neutrino mass generation.

Therefore, the minimal exotic field contents required for non-anomalous $U(1)_X$ needs:
three right-handed neutrinos, two additional ${\bf 5}+\overline{\bf 5}$ pairs beyond the one that provides $H_u$ and $H_d$, 
and two additional charge-$4a$ singlets
beyond $S$. The costs of GUT uniform $U(1)_X$ charge assignment are that the $H_u$ and $H_d$ fields must be completed into SU(5) multiplets, 
which raises a doublet–triplet splitting problem. Light color triplets, whose bare masses are forbidden and the generated masses are correlated 
to the $\mu$-term, would trigger fast proton decay. To avoid such a proton decay problem, the simplest way out is to adopt orbifold GUT, 
where the doublet-triplet splitting among ${\bf 5}_H+\overline{\bf 5}_H$ are realized by orbifold projection after assigning proper boundary 
conditions. An alternative way to avoid the mass correlation between the doublet Higgs and the triplet Higgs is to add additional 
$\lambda 5_H \bar{5}_H \Sigma_{24}$, which however spoil the solution to $\mu$ problem. We will adopt the orbifold GUT approach in the following
 discussions.

For the forbidden of bare $\mu$-term, no additional discrete symmetry are needed, 
since the $U(1)_X$ gauge symmetry alone supplies the selection rule. In particular there is no $\kappa$ parameter, 
and the singlet fermion mass is not a free input of the model. The same condition forbids the Giudice-Masiero operator 
$\tilde{X}^\dagger H_uH_d/M_*$ in the K\"ahler potential for a spurion $\tilde{X}$ neutral under $U(1)_X$.

The $SU(5)$ GUT breaking by boundary conditions is realized with an
$M_4\times S^1/Z_2$ orbifold, which can preserve and break the $SU(5)$ GUT gauge group
at the two fixed point $z=0$ and $z=\pi R$ with Wilson line, respectively~\cite{Hall:2001}.
 MSSM matter contents (plus RH neutrinos) are located on the GUT-preserving fixed point while the 
 three $(\mathbf{5}_{-2}+\bar{\mathbf5}_{-2})$ pairs and singlets are bulk fields. 
 By assigning proper boundary conditions, the triplet Higgses among ${\bf 5}_H+\overline{\bf 5}_H$, doublets within the 
 two $Y(\mathbf5_{-2})+\bar{Y}(\bar{\mathbf5}_{-2})$ pairs, and two singlets can be projected out.

The $SU(5)$-invariant superpotential at the GUT-preserving brane, valid up to the compactification scale, reads
\begin{equation}
 W_{\rm high}=y^U_{ij} {\bf 10}_i {\bf 10}_j 5_H+y^{DE}_{ij} {\bf 10}_i {\bf \bar{5}}_j \bar{5}_H+y^N_{ij} {\bf \bar{5}}_i {\bf 1}_j {5}_H
 +\sum\limits_{a=1}^3\lambda S_a 5_H\bar{5}_H
 +\sum_{a=1}^{3}y_{Y}S_a Y_i\bar{Y}_i~.
\label{eq:whigh}
\end{equation}
 After orbifold projection, the low-energy superpotential of our scenario reads
\begin{equation}
 W_{\rm low}=W_{MSSM}+y^N_{ij} L_i N_j H_u
 +\lambda S H_u H_d
 +\sum_{i=1}^{2}y_{Y}S  \tilde{D}\tilde{D}_i^c .
\label{eq:wlow1}
\end{equation}
Here we identify $S_1\equiv S$. The light matter spectrum consists of the MSSM fields (plus right-handed neutrinos), 
the singlet $S$, and two copies of the extra color-triplet pair $\tilde{D}_i+\tilde{D}_i^c$. 
The projected content is not by itself anomaly-free: the gauge anomalies of the complete $3\times\mathbf{27}$ multiplets must be compensated 
at the orbifold fixed points, or by the projected-out heavy modes.
 The Dirac Yukawa couplings of the right-handed neutrinos is fixed by the observed mass scale, 
$y_\nu\lesssim10^{-12}$ for $m_\nu\lesssim0.1\ev$. So, it can be safely neglected in every soft-breaking expression of appendix~\ref{app:soft}.

\item GUT non-uniform $U(1)_X$ charge assignment:

There are alternative anomaly-free $U(1)_X$ charge assignments. The chiral fields and their charges are given in Table~\ref{tab:fields}.
It can be proven that new chiral superfields $D/E$, three $D+D^c$ and two $E+E^c$, are necessarily present to cancel gauge anomaly.  One can check that the gauge anomaly coefficients 
$(A_{33X},A_{22X},A_{YYX},A_{YXX},
 A_{XXX},A_{{\rm grav}X})$ vanish identically for the choices in the Table~\ref{tab:fields}.
The exotic matter contents required by
exact cancellation also enter the gauge-coupling unification~\cite{Morrissey:2006}.
MSSM matter and $D/E$ fields have odd matter parity, while
$H_d,H_u,S$ have even matter parity. 

Again, no additional discrete symmetry are needed and there is no $\kappa$ parameter. 
Similarly, the Giudice-Masiero operator $\tilde{X}^\dagger H_uH_d/M_*$ in the K\"ahler potential is forbidden for a spurion $\tilde{X}$ neutral 
under $U(1)_X$. Besides, in this scenario, no second singlet are introduced.

Note that, it is obvious that the $U(1)_X$ choices in this scenario are not compatible with ordinary 4D SU(5) GUT, as the matter contents within each GUT multiplet do not have the same $U(1)_X$ quantum number. However, such a low energy spectrum may origin from high-dimensional orbifold GUT with proper chosen boundary conditions, so as that each low energy matter contents surviving orbifold projection from different SU(5) multiplets. If we choose other $U(1)^\prime$ with neutral singlet, for example, $U(1)_{B-L}$, which does not forbid the bare $\mu$-term, the $U(1)^\prime$ charge assignment can be compatible with 4D GUT with the price of additional discrete symmetry.  

\begin{table}[H]
\centering\small
\begin{tabular}{@{}lclrr@{}}
\toprule
Superfield & copies & $SU(3)_c,SU(2)_L$ & $Y$ & $q_X$\\
\midrule
$Q$   &3& $(\mathbf3,\mathbf2)$       &$1/6$  &$-1/18$\\
$u^c$ &3& $(\overline{\mathbf3},\mathbf1)$ &$-2/3$ &$5/9$\\
$d^c$ &3& $(\overline{\mathbf3},\mathbf1)$ &$1/3$  &$5/9$\\
$L$   &3& $(\mathbf1,\mathbf2)$       &$-1/2$ &$1/2$\\
$e^c$ &3& $(\mathbf1,\mathbf1)$       &$1$    &$0$\\
$H_u$ &1& $(\mathbf1,\mathbf2)$       &$1/2$  &$-1/2$\\
$H_d$ &1& $(\mathbf1,\mathbf2)$       &$-1/2$ &$-1/2$\\
$S$   &1& $(\mathbf1,\mathbf1)$       &$0$    &$1$\\
$D$   &3& $(\mathbf3,\mathbf1)$       &$-1/3$ &$-5/9$\\
$D^c$ &3& $(\overline{\mathbf3},\mathbf1)$ &$1/3$ &$-4/9$\\
$E$   &2& $(\mathbf1,\mathbf1)$       &$-1$   &$0$\\
$E^c$ &2& $(\mathbf1,\mathbf1)$       &$1$    &$-1$\\
\bottomrule
\end{tabular}
\caption{Active low-energy chiral content. Copies include three ordinary
families and three $D+D^c$ and two $E+E^c$ vectorlike pairs.}
\label{tab:fields}
\end{table}

The low-energy superpotential in our scenario is given by
\begin{equation}
 W_{\rm low}=y_u QH_u u^c+y_d QH_d d^c+y_e LH_de^c
 +\lambda S H_u\!\cdot\!H_d
 +\sum_{i=1}^{3}y_{D_i}SD_iD_i^c
 +\sum_{a=1}^{2}y_{E_a}SE_aE_a^c .
\label{eq:wlow2}
\end{equation}
In our numerical calculation, we use real parameters, diagonal visible-family
Yukawas and universal $y_D,y_E$ for exotic Yukawa couplings.

The low energy one-loop beta-functions and physical Abelian charge trace are
\begin{equation}
 (b_1,b_2,b_3,b_X)=\Bigl(\tfrac{51}{5},1,0,\tfrac{47}{3}\Bigr),
 \qquad
 B_{(Y,X)}=\sum_i n_i q_iq_i^T
 =\begin{pmatrix}17&-5\\-5&\tfrac{47}{3}\end{pmatrix}.
\label{eq:traces}
\end{equation}
Here $q_i=(Y_i,q_{X,i})^T$, $n_i$ includes family, color and weak
multiplicities.

\eit

After $H_u,H_d,S$ acquire the Vacuum Expectation Values (VEV) of form $\langle H_{d,u}^0\rangle=v_{d,u}/\sqrt2$ and
$\langle S\rangle=v_s/\sqrt2$, the $U(1)_X$ and electroweak symmetry are broken with
\begin{equation}
 v^2=v_d^2+v_u^2,\qquad
 \tan\beta=v_u/v_d,\qquad
 v_X^2=v^2+v_S^2\simeq v_S^2,
\label{eq:vev-conventions}
\end{equation}
for $v\simeq246\GeV$. 

After the spontaneously broken of $U(1)_X$ by the VEV of $S$, effective $\mu$-term in Eq.~\eqref{eq:wlow1} (Eq.~\eqref{eq:wlow2} ) and the mixing between singlino and $U(1)_X$ gaugino can be generated as
\begin{equation}
 \mu_{\rm eff}=\lambda s=\frac{\lambda v_s}{\sqrt2},
 \qquad
 m_{\widetilde S}=g_X q_S v_s=\frac{\sqrt2\,g_Xq_S}{\lambda}\mu_{\rm eff}.
\label{eq:mueff}
\end{equation}
Equation~\eqref{eq:mueff} is totally different from a
neutral-singlet construction. In our case, $S$ is charged under $U(1)_X$ and no cubic
superpotential term is allowed, the VEV of $S$ generates the Higgsino mass and
sets the $U(1)_X$ breaking scale at the same time. The ratio
$m_{\widetilde S}/\mu$ is fixed by $\sqrt2g_Xq_S/\lambda$ rather than by an
independent coupling. Obtaining a viable electroweak vacuum, and the
magnitude of $v_s$ itself, requires proper soft SUSY breaking
parameters~\cite{Cvetic:1997,Erler:2002,Han:2004,Hundi:2009,Bae:2019,Cohen:2008}.
One quantitative consequence of the charge assignment should be stated here.
The $U(1)_X$ $D$-term vanishes along $q_Sv_s^2+q_{H_u}v_u^2+q_{H_d}v_d^2=0$,
i.e.\ $v_s^2=(v_u^2+v_d^2)/2$ for $q_S=0.4$ and $q_{H_u}=q_{H_d}=-0.2$. So, a
vacuum close to that direction has
$\mu_{\rm eff}=\lambda v_s/\sqrt2\simeq\lambda v/2\simeq123\lambda\GeV$:
a TeV-scale $\mu_{\rm eff}$ then needs large value of $\lambda$, or a soft-mass
hierarchy that lifts $v_s$ off the $D$-flat direction. 
Besides, because of the last term in Eq.~\eqref{eq:wlow1} (or the last two terms in Eq.~\eqref{eq:wlow2}), 
which are allowed by the $U(1)_X$ charge assignment, mass terms for exotic $Y_i,\bar{Y}_i$ (or $D,D^c$ and $E,E^c$) 
will also be generated after the spontaneously broken of $U(1)_X$. It can be seen that gauge symmetry and parity also 
permit $Dd^c$ and $Ee^c$ bilinears, which we set to vanish in Eq.~\eqref{eq:wlow2}~\footnote{For non-vanishing proper $Dd^c$ 
and $Ee^c$ bilinears, top-seesaw type scenarios can be realized.}. This restricted interaction can affect exotic decays, 
which will be briefly noted in the last section.

\subsection{The neutralino mixing matrix}
\label{sec:matrix}
The inelastic interpretation of LZ high recoil event prefers the two lightest
neutral states to be the nearly degenerate Higgsino pair, which must be
checked against the full six-state spectrum rather than assumed.
For real parameters, take the gauge-eigenstate basis
\begin{equation}
 \psi^0=(\widetilde W^0,\widetilde B,\widetilde B_X,
              \widetilde H_u^0,\widetilde H_d^0,\widetilde S)^T,\qquad
 \mathcal L_{\rm mass}=-\frac12(\psi^0)^T\mathcal M_6\psi^0+\mathrm{h.c.}
\label{eq:matrix-basis}
\end{equation}
There are exactly six gauge-eigenstates: the two electroweak gauginos, the
$U(1)_X$ gaugino, the two Higgsinos and the single singlino. Because $S$
carries $U(1)_X$ charge, $\widetilde B_X$ does not decouple, and the
neutralino matrix does not split into a reduced block plus a separate
sector. With $s_W=\sin\theta_W$, $c_W=\cos\theta_W$,
$s_\beta=\sin\beta$, $c_\beta=\cos\beta$ and the VEV
convention of eq.~\eqref{eq:vev-conventions},
\begin{equation}
 \mathcal M_6=\begin{pmatrix}
 M_2&0&0&-s_\beta c_Wm_Z&c_\beta c_Wm_Z&0\\
 0&M_1&M_{1X}&s_\beta s_Wm_Z&-c_\beta s_Wm_Z&0\\
 0&M_{1X}&M_{XX}&\xi_uv_u&\xi_dv_d&m_{\widetilde S}\\
 -s_\beta c_Wm_Z&s_\beta s_Wm_Z&\xi_uv_u&0&-\mu&-\lambda v_d/\sqrt2\\
 c_\beta c_Wm_Z&-c_\beta s_Wm_Z&\xi_dv_d&-\mu&0&-\lambda v_u/\sqrt2\\
 0&0&m_{\widetilde S}&-\lambda v_d/\sqrt2&-\lambda v_u/\sqrt2&0
 \end{pmatrix},
\label{eq:matrix}
\end{equation}
with
\begin{equation}
 m_{\widetilde S}=g_Xq_Sv_s,\qquad
 \xi_u=g_Xq_{H_u}+\frac{g_{YX}}{2},\qquad
 \xi_d=g_Xq_{H_d}-\frac{g_{YX}}{2},
\label{eq:xi-def}
\end{equation}
where $g_{YX}$ is the Abelian kinetic-mixing entry and $M_{1X}$ the
off-diagonal gaugino mass. The superpotential term in
eq.~\eqref{eq:discrete-selection} gives the $\lambda v_{u,d}/\sqrt2$
entries, and the $U(1)_X$ charges of the Higgs doublets give the additional
$\xi_{u,d}v_{u,d}$ entries, which have no counterpart in a
neutral-singlet construction. The $(\widetilde S,\widetilde B_X)$ block
\begin{equation}
 \begin{pmatrix}0&m_{\widetilde S}\\
 m_{\widetilde S}&M_{XX}\end{pmatrix},
 \qquad
 \lambda_{X,\pm}=\frac{M_{XX}\pm\sqrt{M_{XX}^2+4m_{\widetilde S}^2}}{2},
\label{eq:xs-block}
\end{equation}
gives the two eigenstates. For $M_{XX}\ll m_{\widetilde S}$ these are approximately $\pm m_{\widetilde S}$.
The sign of the parameters $M_1$, $M_2$, $M_{XX}$ and $M_{1X}$ must be retained. Keeping only their magnitudes changes the interference 
that controls the splitting.
 We adopt the general form with gauge kinetic mixing, and therefore a non-zero $g_{YX}$ in the previous neutralino matrix.

For a real symmetric matrix, diagonalization gives signed
eigenvalues and the physical masses are their absolute values. In our setting, the lightest state must be predominantly Higgsino.

\subsection{The neutral Higgsino mass splitting}
\label{sec:splitting}

Define the physical splitting of the Higgsino-like light pair by
\begin{equation}
 \Delta m_\chi=m_{\chi_2}-m_{\chi_1}\ge0.
\label{eq:splitting-definition}
\end{equation}
This is distinct from the messenger parameter $\delta_{\rm mess}$.
To exhibit its origin, reorder the six states of eq.~\eqref{eq:matrix} as
$(\widetilde H_d^0,\widetilde H_u^0;\widetilde B,\widetilde W^0,
\widetilde S,\widetilde B_X)$ and write the matrix in block form,
\begin{equation}
 \mathcal M_6=\begin{pmatrix}
 \mathcal M_H^{(0)}&\mathcal V\\
 \mathcal V^T&\mathcal M_R
 \end{pmatrix},\qquad
 \mathcal M_H^{(0)}=\begin{pmatrix}0&-\mu\\-\mu&0\end{pmatrix},
\label{eq:block-form}
\end{equation}
with the Higgsino-to-heavy mixing
\begin{equation}
 \mathcal V=\begin{pmatrix}
 -c_\beta s_Wm_Z&c_\beta c_Wm_Z&-\lambda v_u/\sqrt2&\xi_dv_d\\
 s_\beta s_Wm_Z&-s_\beta c_Wm_Z&-\lambda v_d/\sqrt2&\xi_uv_u
 \end{pmatrix}
\label{eq:heavy-light-block}
\end{equation}
and the heavy $4\times4$ block $\mathcal M_R$ built from the parameters
$(M_1,M_2,m_{\widetilde S},M_{XX},M_{1X})$ as in eq.~\eqref{eq:matrix}. 

When $|\mu|$ is small compared with the eigenvalues of $\mathcal M_R$, the
leading correction to the two-Higgsino block is
$-\mathcal V\mathcal M_R^{-1}\mathcal V^T$, and the pair splitting is the
absolute value of its trace. For vanishing kinetic mixing, the splitting can be evaluated
by the following form
\begin{equation}
 \Delta m_\chi\simeq\left|
 m_Z^2\left(\frac{s_W^2}{M_1}+\frac{c_W^2}{M_2}\right)
 +\frac{\lambda^2v^2\sin2\beta}{2\mu}
 -\frac{\lambda^4M_{XX}v^2}{4g_X^2q_S^2\mu^2}
 +\frac{M_{1X}^{2}\lambda^{2}v^{2}}{2M_1g_X^{2}q_S^{2}v_s^{2}}\right|,
\label{eq:singlet-splitting}
\end{equation}
with $v^2=v_u^2+v_d^2$. The first term is the familiar electroweak gaugino
contribution. The second is produced by the $\widetilde B_X$--Higgsino entries
$\xi_{u,d}v_{u,d}$ of eq.~\eqref{eq:heavy-light-block}, inserted through the
$\widetilde S$--$\widetilde B_X$ mixing and reduced with the charge sum rule
$q_S+q_{H_u}+q_{H_d}=0$. It plays the role of the Higgsino-singlino term of
singlet extensions~\cite{HiggsinoNMSSM}, but it needs no superpotential
coupling. The third is the $U(1)_X$ gaugino correction, which mixes the
singlino with $\widetilde B_X$ through $m_{\widetilde S}$, and the fourth
arises from the off-diagonal gaugino mass $M_{1X}$. The $1/q_S^2$ factors in
the last two terms follow from $m_{\widetilde S}=g_Xq_Sv_s=\sqrt2g_Xq_S\mu/\lambda$.
All four terms are added with their signs before the absolute value is taken,
so they can interfere constructively or destructively.

Equivalently, defining
\begin{equation}
 C_g=m_Z^2\left(\frac{s_W^2}{M_1}+\frac{c_W^2}{M_2}\right),\qquad
 C_X=\frac{\lambda^2v^2}{2\mu}
     \left(\sin2\beta-\frac{\lambda^2M_{XX}}{2g_X^2q_S^2\mu}\right),
\label{eq:splitting-coefficients}
\end{equation}
one has $\Delta m_\chi\simeq|C_g+C_X|$ for $M_{1X}=0$, to be compared with $|C_g+C_s|$
of a neutral-singlet model~\cite{EllwangerReview} with
$C_s=\lambda^2v^2/(2m_{\widetilde S})$.
There are two structural differences between the two scenarios. First, eq.~\eqref{eq:mueff} ties the
singlino scale to $\mu$,
\begin{equation}
 \frac{m_{\widetilde S}}{\mu}=\frac{\sqrt2g_Xq_S}{\lambda},
\label{eq:singlino-ratio}
\end{equation}
so the denominator of the singlino term is fixed by the ratio of a gauge
coupling to a Yukawa coupling rather than by an independent cubic
coupling. Second, the kinetic mixing does not enter
eq.~\eqref{eq:singlet-splitting} at all: its contribution to the trace
vanishes identically, because only the combination
$\xi_d+\xi_u=g_X(q_{H_u}+q_{H_d})$ appears and that is $g_{YX}$-independent. The off-diagonal gaugino
mass $M_{1X}$ enter quadratically through
$\mathcal M_R^{-1}$. Equation~\eqref{eq:singlet-splitting} is the leading expression for $\Delta m_\chi$ and not a
spectrum calculation. It is reliable when every eigenvalue of
$\mathcal M_R$ is well separated from $\mu$. 

The valid range of eq.~\eqref{eq:singlet-splitting} is not solely controlled by
$m_{\widetilde S}$. The lighter eigenvalue of the
$(\widetilde S,\widetilde B_X)$ block of eq.~\eqref{eq:xs-block} is
\begin{equation}
 \lambda_{X,-}=\frac{M_{XX}-\sqrt{M_{XX}^2+4m_{\widetilde S}^2}}{2}
 \;\longrightarrow\;
 \begin{cases}
 -\dfrac{m_{\widetilde S}^{2}}{M_{XX}}, & M_{XX}\gg m_{\widetilde S},\\[2mm]
 -m_{\widetilde S}, & M_{XX}\ll m_{\widetilde S},
 \end{cases}
\label{eq:light-eigenvalue}
\end{equation}
so a heavy singlino requires
\begin{equation}
 \frac{m_{\widetilde S}^{2}}{M_{XX}}\gg|\mu|
 \qquad\Longleftrightarrow\qquad
 M_{XX}\ll\frac{2g_X^{2}q_S^{2}\mu}{\lambda^{2}},
\label{eq:mxx-validity}
\end{equation}
and not merely $m_{\widetilde S}\gg|\mu|$. When $M_{XX}$ violates
eq.~\eqref{eq:mxx-validity}, the lightest state is singlino-dominated. So, the previous
expansion fails and one needs to diagonalize the full six-state matrix of eq.~\eqref{eq:matrix}.

Two numerical comparisons confirm this. With
$g_X=1.0$, i.e.\ $g_Xq_S=0.4$ and $m_{\widetilde S}=\sqrt2g_Xq_S\mu/\lambda=6.2\TeV$, $\tan\beta=20$, $M_1=50\TeV$, $M_2=120\TeV$, $\lambda=0.1$ and
$\mu=1.1\TeV$, the previous form reproduces the exact six-state
Higgsino-pair splitting to between $1\%$ and $6\%$ for
$M_{XX}\le1\TeV$ (where $\min|\lambda_{X,\pm}|/\mu\ge5.2$), degrading to
$15\%$ at $M_{XX}=4\TeV$. At the smaller mass separations
$M_1=3\TeV$, $M_2=6\TeV$, $M_{XX}=0.3\TeV$, it agrees to better than
$11\%$ for $\lambda\le0.4$ (where the lightest state has a Higgsino
probability above $0.999$), while for $\lambda\ge0.6$ it underestimates
the splitting by more than a factor of two and the lightest state is no
longer Higgsino-dominated. In the decoupled limit, the MSSM result is recovered,
\begin{equation}
 \Delta m_\chi^{\rm gaugino}\simeq
 m_Z^2\left|\frac{s_W^2}{M_1}+\frac{c_W^2}{M_2}\right|.
\label{eq:gaugino-split}
\end{equation}
Opposite signs of $M_1$ and $M_2$ can suppress this term~\cite{DuWang},
but the singlino terms of eq.~\eqref{eq:singlet-splitting} are not
suppressed by that choice and must be controlled as well: at
$\lambda=0.6$, $\mu=1.1\TeV$ and $\tan\beta=20$, the second term is
about $1\GeV$ for any $M_{XX}$, while the third, which grows with
$M_{XX}$, reaches $10\GeV$ at the validity bound of
eq.~\eqref{eq:mxx-validity}. Either is roughly three orders of magnitude
above the sub-MeV scale that inelastic direct detection requires.
Reaching such a splitting in this model therefore requires either
$\lambda\lesssim0.01$ or a cancellation against the gaugino term at the
$10^{-3}$ level.

Written in term of the singlino mass, the expression 
\begin{equation}
 \frac{\lambda^{2}v^{2}\sin2\beta}{2\mu}
 =\frac{\lambda g_Xq_S}{\sqrt2}\,
 \frac{v^{2}\sin2\beta}{m_{\widetilde S}},
\label{eq:singlino-1overmsusy}
\end{equation}
can be compared with $C_s=\lambda^{2}v^{2}/(2m_{\widetilde S})$ for a
neutral-singlet model. Obviously, the dependence on $m_{\widetilde S}$ agrees, while the
numerators differ: it is $\lambda g_Xq_S\sin2\beta/\sqrt2$ here instead of
$\lambda^{2}/2$ for neutral-singlet model. In fact, in the charged-singlet model, the singlino mass originates from both the Abelian $D$ term and $U(1)_X$ gaugino mass,
rather than from the cubic superpotential coupling of the form $S^3$. 

The charged--neutral ordering is a separate question. At tree level, the
chargino is lighter than the lightest Higgsino only in a restricted corner for MSSM.
Writing
\begin{equation}
 X=\frac{c_W^2}{M_2}+\frac{s_W^2}{M_1},\qquad
 Y=\frac{c_W^2}{M_2}-\frac{s_W^2}{M_1},
\label{eq:xy-def}
\end{equation}
the leading tree-level results in the Higgsino limit
$m_Z\ll|\mu|\ll|M_{1,2}|$, for $\mu>0$, are
$\Delta m_0^{\rm tree}=m_Z^2|X|$ together with
\begin{equation}
 m_{\chi_1^\pm}-m_{\chi_1^0}
 =\frac{m_Z^2}{2}\Bigl(|X|-\sin2\beta\,Y\Bigr).
\label{eq:chargino-tree}
\end{equation}
So the chargino is lighter when $\sin2\beta\,Y>|X|$. In our MSSM plus $U(1)_X$
model, the singlino-Higgsino mixing through the off-diagonal entries
$\xi_{u,d}v_{u,d}$ of eq.~\eqref{eq:heavy-light-block} introduces an
additional tree-level contribution. To leading order in
$m_Z/|\mu|$, $m_Z/|M_{1,2}|$ and $\lambda v/m_{\widetilde S}$, with
$m_{\widetilde S}=\sqrt2 g_X q_S\mu/\lambda$ from eq.~\eqref{eq:mueff} and
$M_{1X}=0$, the lightest neutralino mass receives a shift
$-\delta_c$ relative to the MSSM value implicit in
eq.~\eqref{eq:chargino-tree}, giving
\begin{equation}
\begin{split}
 m_{\chi_1^\pm}-m_{\chi_1^0}
 &=\frac{m_Z^2}{2}\Bigl(|X|-\sin2\beta\,Y\Bigr)+\delta_c,\\[2mm]
 \delta_c&=\frac{\lambda^2v^2}{8g_X^2\mu^2}
 \biggl[-0.2g_Xg_{YX}\mu\cos2\beta-\Bigl(2g_X^2\mu-\lambda^2M_{XX}\Bigr)(1+\sin2\beta)\biggr].
\end{split}
\label{eq:chargino-u1x}
\end{equation}
The dominant piece of $\delta_c$ is
$-\lambda^2v^2(1+\sin2\beta)/(4\mu)$, which is negative and of order
$150\MeV$ at $\lambda=0.1$, $\mu=1.1\TeV$. It arises from the
$\widetilde S$--$\widetilde H$ mixing via $\xi_{u,d}v_{u,d}$ and lowers
$m_{\chi_1^0}$ relative to the chargino. The $M_{XX}$ term partially
compensates when $M_{XX}$ is large, and the $g_{YX}$ term encodes the
kinetic-mixing dependence ($\cos2\beta<0$ for $\tan\beta>1$ makes it
positive). Equation~\eqref{eq:chargino-u1x} is valid when every
eigenvalue of $\mathcal M_R$ is well separated from $\mu$ and the
lightest state remains Higgsino-like. For $\lambda\gtrsim0.3$ or small
$M_{XX}$, one must diagonalise the full six-state matrix. The one-loop electroweak correction
raises the charged state by about $340$ to $353\MeV$ in the Higgsino
limit~\cite{IbeHiggsino}. The $U(1)_X$ correction $\delta_c$ of eq.~\eqref{eq:chargino-u1x}
shifts the tree-level splitting by ${\sim}150\MeV$ at $\lambda=0.1$,
comparable to the one-loop electroweak contribution. For
$\lambda\gtrsim0.3$, it exceeds the loop term and can reverse the
ordering, excluding those parameter points from the inelastic
interpretation. Within the viable range ($\lambda\lesssim0.15$) the
neutral Higgsino remains the lightest $R$-odd state. The same interpretation predicts a chargino a few hundred $\MeV$ above the neutral states, and therefore a
short charged track at future colliders~\cite{CheungLZ}. 

\section{Deflected anomaly mediation and the messenger sector}
\label{sec:damsb}
In our MSSM plus $U(1)_X$ scenario (with an additional singlet sector), the soft SUSY breaking parameters are put in by hand. 
Such soft SUSY parameter inputs should be determined by proper SUSY breaking mechanism, so as that only few free parameters are needed. 
Depending on the way  the visible sector $'feels'$ the SUSY breaking effects in the hidden sector, the SUSY breaking mechanisms can be 
classified into gravity mediation~\cite{SUGRA}, gauge mediation~\cite{GMSB}, anomaly mediation~\cite{Randall:1998uk,Giudice:1998xp} scenarios, 
etc. Especially, the predictive AMSB and GMSB mechanisms will not cause flavor and CP problems that bothers gravity/modulus mediation models. 
We know that the inelastic interpretation of the LZ high-recoil event prefer a nearly degenerate Higgsino pair, 
hence a Higgsino LSP. However, the GMSB mechanism is not viable, as it always predicts a very light gravitino LSP, which would act as the 
dark-matter candidate instead. So, we would like to adopt the predictive AMSB type mechanism to act the UV completion of our model.

Minimal AMSB, which is determined solely by the parameter $m_{3/2}$,
is insensitive to its UV theory~\cite{AMSB:RGE} and predicts a flavor conservation soft SUSY breaking spectrum. 
Unfortunately, negative slepton squared masses are predicted and the minimal scenario must be extended. 
Although there are many possible ways to tackle such tachyonic slepton masses problem, the most elegant solution from aesthetical point of view 
is the deflected AMSB~(dAMSB)~\cite{Pomarol:1999ie,Rattazzi:1999qg,Okada:2002mv,Okada:2012nr,Wang:2015apa} scenario, 
which adopts non-trivially an additional messenger sector to deflect the AMSB trajectory and push the negative slepton squared masses 
to positive values by additional gauge mediation contributions.  
After the renormalization group equation~(RGE) evolution below the messenger thresholds, 
the anomaly mediation contributions will receive threshold corrections after integrating out the messenger fields. 
The soft SUSY breaking parameters in deflected AMSB are given by mixed anomaly mediation and gauge mediation contributions 
in addition to pure AMSB and GMSB ones.

\subsection{Soft SUSY breaking parameters in deflected AMSB}
The form of the superpotential upon the messenger threshold includes the messenger sectors and superpotential $W(\tilde{X})$ 
for pseudo-moduli $\tilde{X}$ with
\beqa
W\supseteq W_{low}+\la_{\tilde{X}} \tilde{X} \bar\Psi_{\overline{\mathbf5}}\Psi_\mathbf5 + W(\tilde{X})~.
\eeqa
 The pseudo-moduli superfield $\tilde{X}$  will receive F-term VEV from anomaly mediation and 
 has the form $<\tilde{X}>=M+\theta^2 F_{\tilde{X}}$.
The deflection parameter, which characterize the deviation from the ordinary AMSB trajectory, depends on the form of the pseudo-moduli 
superpotential $W(\tilde{X})$. Such superpotential can be fairly generic and lead to a deflection parameter $'d'$ of either sign given by
\beqa
d F_\phi\equiv \f{F_{\tilde{X}}}{{\tilde{X}}}-F_\phi~.
\eeqa
Here $F_\phi$ is the F-term VEV for the compensator superfield $\phi$, with $<\phi>=1+\theta^2 F_\phi$.
 The wavefunction renormalization approach~\cite{wavefunction:hep-ph/9706540,chacko} can be used in deflected AMSB~\cite{Fei:1602.01699,Fei:1703.10894} to keep track of supersymmetry-breaking effects with
a manifestly supersymmetric formalism and obtain the soft SUSY breaking spectrum. In deflected AMSB, the messenger threshold $M_{mess}$ and 
the RGE scale $\mu$ in the wavefunction superfield and gauge kinetic superfield can be replaced by the following combinations involving 
the spurious chiral fields $\tilde{X}$
\beqa
 M_{mess}\ra \sqrt{\f{\tilde{X}^\da \tilde{X}}{\phi^\da\phi}}~,~~~~~\mu\ra \f{\mu}{\sqrt{\phi^\da\phi}}~,
\eeqa
which, after substituting the F-term VEV of $\tilde{X}$ and $\phi$, can obtain the soft SUSY breaking parameters.

The messenger sector consists of $N$ vector-like complete $SU(5)$ 
$\mathbf5+\overline{\mathbf5}$ pairs,
\begin{equation}
 \Psi(\mathbf5)=(\mathbf3,\mathbf1)_{Y=-1/3}\oplus(\mathbf1,\mathbf2)_{Y=+1/2},
 \qquad
 \bar\Psi(\overline{\mathbf5})=(\overline{\mathbf3},\mathbf1)_{Y=+1/3}
                     \oplus(\mathbf1,\mathbf2)_{Y=-1/2},
 \label{eq:soft-mess-content}
\end{equation}
which will not re-introduce gauge anomaly for $SU(3)_c\times SU(2)_L\tm U(1)_Y\tm U(1)_X$.
The $U(1)_X$ charge assignment of the messengers is taken universal within each complete multiplet, 
$q_X(\mathbf5_M)=+2q$ and $q_X(\overline{\mathbf5}_M)=-2q$, so that the pair is vectorlike under both the standard-model group and $U(1)_X$. 
The factor two is not optional, since the Kahler portals of eq.~\eqref{eq:messenger-portal} require $q_M=\pm 2q$ 
(section~\ref{sec:messenger-link}). The messenger contribution to $b_i$ is then the universal shift $\Delta b_i=N$ together 
with $\Delta b_X=2N(3q_M^2+2q_M^2)$.

In a ${\bf5}+{\bf\bar5}$ pair, the two messenger scalars have masses $\la_{\tilde{X}}^2 M^2\pm \la_{\tilde{X}} F$ and the
messenger Dirac fermion has mass $\la_{\tilde{X}} M$. We define the messenger scale at $M_{mess}\equiv \la_{\tilde{X}} M$ So, the lightest messenger is normally the
lighter of the two scalars. We assign the messenger superfields even
$R$-parity, a convention that leaves the mass term and the gauge
interactions invariant and keeps matter parity anomaly-free. With this
assignment, the messenger scalar is $R$-even. For $R$-even lightest messenger scalar, if it is not stable, the $R$-parity requires an
{even} number of MSSM $R$-odd states in its possible decay final states. For example, the decay $\varphi\to\tilde f\bar f$ is forbidden, 
while all SM and two-sfermion final states are allowed.

For $N$ messenger pairs, for the messenger content given in
eq.~\eqref{eq:soft-mess-content}, the gaugino masses for 
deflected-AMSB are given by
\begin{equation}
 M_i(M_{\rm mess})=-\frac{F_\phi g_i^2(M_{\rm mess})}{16\pi^2}
                  (b_i-\Delta b_N  d),
 \label{eq:gaugino}
\end{equation}
with $\Delta b_N$ the new contributions of messengers to each beta functions.
The gauge beta function coefficients for $U(1)_Y\tm SU(2)_L\tm SU(3)_c\tm U(1)_X$ below the messenger threshold are given by 
$(b_1,b_2,b_3,b_X)=({37}/{5},1,-1,1.28)$
for the GUT uniform $U(1)_X$ charge assignment \eqref{eq:gut-uniform-charges} with the projected light spectrum of eq.~\eqref{eq:wlow1}, 
and by $b_i=(51/5,1,0,47/3)$ for the GUT non-uniform charge assignment of Table~\ref{tab:fields}.
In both cases the extra color-triplet supermultiplets are active low-energy contents, and their contributions are already 
included in $b_i$.  For a fixed effective field content and fixed delfection parameter $d$,
$F_\phi$ sets the overall gaugino scale while $d$ changes the ratios.
At $d=0$ the deflection is absent, giving a pure AMSB contribution. 

We present the analytical expressions of the soft SUSY breaking parameters for our $U(1)_X$ extended MSSM model 
in appendix~\ref{app:soft}. The coefficients and the running must 
be evaluated in the effective theory appropriate to each threshold. 
Such analytical expressions that predicted by deflected AMSB are valid at the messenger scale. 
To obtain the low energy soft SUSY breaking spectrum, we need to RGE evolve down from the messenger scale to the electroweak scale, 
with proper matching across each threshold, such as the $U(1)_X$ breaking threshold. 

A Higgsino candidate requires $|\mu|$ below the physical bino, wino,
singlino and $U(1)_X$ gaugino eigenmasses, with the mixing given by
eq.~\eqref{eq:matrix}. In a well-decoupled limit this suggests
$|M_1|,|M_2|,|m_{\widetilde S}|,|M_{XX}|>|\mu|$, but near-degeneracies
and mixing make a full diagonalization necessary. The additional scalar
sector must also be non-tachyonic. For the messenger pair this requires the messenger scalars to be
non-tachyonic, which in the present setup amounts to
$(1+d)F_\phi<M_{\rm mess}$: with $\langle {\tilde{X}}\rangle=M_{\rm mess}/\la_{\tilde{X}}$ and
$F_{\tilde{X}}=(1+d)\langle {\tilde{X}}\rangle F_\phi$, the condition
$\lambda_{\tilde{X}}\langle {\tilde{X}}\rangle^2>|F_{\tilde{X}}|$ reduces to this form and is independent of
$\lambda_{\tilde{X}}$. 

\begin{figure}[tbp]
 \centering
 \includegraphics[width=\textwidth]{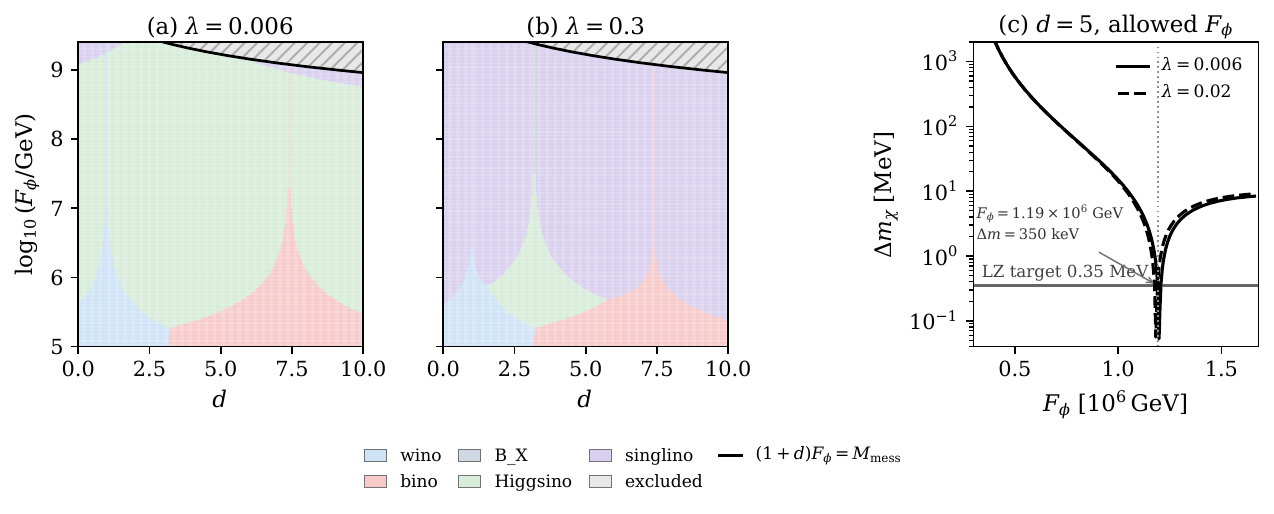}
 \caption{LSP composition and Higgsino-pair splitting in the deflected-AMSB
 parameter space, for $N=1$, $M_{\rm mess}=10^{10}\GeV$, $\tan\beta=20$,
 $\mu=1.1\TeV$, $g_X(M_{\rm mess})=1.4$, messenger charge
 $q_X(\mathbf5_M)=\pm 2q$ and the uniform coefficients $(37/5,1,-1,1.28)$, with
 $g_{YX}=M_{1X}=0$. Panels (a) and (b) show which
 neutralino is the LSP in the $d$--$F_\phi$ plane for $\lambda=0.006$ and
 $\lambda=0.3$. The narrow structures near $d=1$ and $d=7.4$ are the strips in
 which $M_2$ and $M_1$ change sign.
The hatched region is excluded by the messenger
non-tachyonic condition, i.e.\ it is the part of the plane where
$(1+d)F_\phi>M_{\rm mess}$. Panel (c)
gives $\Delta m_\chi$ versus $F_\phi$ at $d=5$, where the two gaugino contributions to the splitting carry opposite signs. 
The horizontal line is the LZ target $0.35\MeV$, which the curve crosses twice. 
The filled point marks the crossing at $F_\phi=1.19\times10^{6}\GeV$, where $\Delta m_\chi=350\keV$.}
 \label{fig:lsp-scan}
\end{figure}

The composition of the lightest neutralino in this parameter space is shown
in fig.~\ref{fig:lsp-scan}. We take $N=1$, $M_{\rm mess}=10^{10}\GeV$,
$\tan\beta=20$, $\mu=1.1\TeV$, and $g_X(M_{\rm mess})=1.4$ (which gives \ $g_X\simeq1.1$ at
the TeV scale) that chosen to keep the
Landau pole above the GUT scale ($\Lambda\simeq2.6\times10^{20}\GeV$), the deflection parameter
$d\in[0,10]$ and $F_\phi\in[10^5,10^9]\GeV$. We adopt the one-loop RGE equations for
$g_i$ and $M_i$ with the coefficients $(37/5,1,-1,1.28)$ for GUT uniform case, and
diagonalize the six-state matrix \eqref{eq:matrix} with $g_{YX}=M_{1X}=0$,
both of which vanish identically for the uniform content.

Within the region allowed by $(1+d)F_\phi<M_{\rm mess}$, above a
$d$-dependent threshold, the LSP is Higgsino-like over most of it. 
The value of $\lambda$ decides whether this Higgsino region is available
at all. For $\lambda=0.3$ (panel b), the singlino is the LSP over about
three quarters of the region allowed by the non-tachyonic
condition, because eq.~\eqref{eq:mxx-validity} requires
$M_{XX}\ll2g_X^2q_S^2\mu/\lambda^2\simeq5.1\TeV$ at this point, whereas
$|M_{XX}|=(F_\phi/16\pi^2)g_X^2|0.4Nd-1.28|$ grows with $F_\phi$. The Higgsino window therefore
survives only at low $F_\phi$: with $M_{\rm mess}=10^{10}\GeV$ the non-tachyonic bound is satisfied over the whole plotted range, so the window is closed by eq.~\eqref{eq:mxx-validity} alone, and the vanishing of $M_{XX}$ at $d\simeq3.2$ lies inside the allowed region, where it produces the narrow Higgsino strip of panel (b). For
$\lambda=0.006$ (panel a), the threshold moves up to
$2g_X^2q_S^2\mu/\lambda^2\simeq1.3\times10^7\GeV$. So, the singlino is heavy over
almost the whole plot, and the Higgsino region fills about three quarters of the allowed range.

The same scan fixes the splitting. Panel (c) evaluates $\Delta m_\chi$ at
$d=5$ as a function of $F_\phi$. With such deflection, the two gaugino
contributions to $C_g$ carry opposite signs. So $\Delta m_\chi$ falls
steeply with $F_\phi$, passes through zero and rises again, and the LZ
window is crossed on the way down and again on the way up.
For $\lambda=0.006$, the crossing $\Delta m_\chi=0.35\MeV$ occurs at
$F_\phi=1.19\times10^{6}\GeV$, where $m_{\chi_1^0}=1100\GeV$ and the Higgsino
fraction is $0.9999$. The messenger non-tachyonic bound is also satisfied there, so this point is a genuine inelastic benchmark of the
model rather than an extrapolation. The window is nevertheless narrow: varying $F_\phi$ by $\pm0.1\%$
moves $\Delta m_\chi$ between $0.30$ and $0.41\MeV$, while displacing $d$ by
$\pm0.1$ at fixed $F_\phi$ raises it to $35$--$38\MeV$, two orders of
magnitude above the required value. The sub-MeV value is therefore a residue of
cancellations rather than a generic choice: at $d=5$ the two gaugino terms
$m_Z^2s_W^2/M_1$ and $m_Z^2c_W^2/M_2$, each of order $0.5\GeV$ and of opposite
sign, already cancel to $3\%$, and the remainder is cancelled in turn by the
higher-order mixing corrections retained only in the full diagonalization. Keeping $\Delta m_\chi$ inside the LZ interval $340$--$360\keV$ would then require $|\delta F_\phi/F_\phi|\lesssim2\times10^{-4}$ and $|\delta d|\lesssim3\times10^{-5}$. For $\lambda\gtrsim0.01$, the singlino term alone exceeds $0.35\MeV$,
and for $\lambda\gtrsim0.03$ the splitting exceeds $10\MeV$ everywhere in the
panel. A point that satisfies both the Higgsino-LSP and the $350\keV$
requirements therefore needs small $\lambda$, a deflection near $d=5$ where
the gaugino terms cancel, and either large $F_\phi$ or small $|\mu|$. Within
these conditions, a Higgsino LSP is generic rather than tuned.

The $500\GeV$, $1.1\TeV$ and $100\TeV$ regimes below correspond to
different choices of $F_\phi$, deflection parameters and singlet inputs.

\subsection{From mediation thresholds to late messenger decays}
\label{sec:messenger-link}

The messengers that deflect the soft spectrum may also contain a
long-lived state whose decay occurs after Higgsino freeze-out. The
mediation threshold $M_{\rm mess}$ and the physical mass of that state
need not coincide once messenger-sector splittings are included.
Its initial abundance, lifetime and mean LSP multiplicity are additional
inputs. Decay operators and
any messenger-number breaking must be specified before those inputs
can be predicted.

We therefore use the messenger interpretation as a cosmological
hypothesis and impose the final-yield conditions that it must satisfy.
A small late messenger source can replenish the under-abundance of light Higgsino dark matter, whereas a sufficiently abundant, 
long-lived messenger source can release enough entropy
to dilute a heavy one. Direct and cascade production
of Higgsinos can possibly spoil the dilution, unless the source decay essentially carries
no Higgsino at all. Both effects are kept in the same equation in the
next section, where the two branches are separated. 

The relevant messenger lifetimes depend on which interactions connect
the messenger sector to the visible sector. In the present model there
are no direct messenger--matter Yukawa couplings in the superpotential:
the messenger fields carry $U(1)_X$ charge but do not mix with MSSM
matter multiplets. Consequently the tree-level two-body decay
$\phi\to{\rm MSSM}$ mediated by a messenger--matter Yukawa coupling $y$,
with width $\Gamma_y=y^2M_{\rm mess}/(8\pi)$, is absent.

Pure gauge interactions alone do not induce a single-messenger
decay: the renormalizable Lagrangian conserves a global messenger number
under which $\Phi\to e^{i\alpha}\Phi$, $\bar\Phi\to e^{-i\alpha}\bar\Phi$
while all MSSM fields are neutral. As emphasised in
ref.~\cite{Dimopoulos:1996nbb}, this symmetry makes the lightest
messenger absolutely stable at the renormalizable level. A decay $\varphi\to{\rm MSSM}$ requires an
explicit messenger-number-violating operator.

The complete set of dimension-five operators that violate messenger number by
one unit is catalogued in~\cite{Dimopoulos:1996nbb}, suppressed by the Planck mass. In $SU(5)$ notation, with
$5_M,\bar5_M$ denoting the messenger superfields and
$\bar5_F,10_F$ the ordinary-family matter superfields, these are
\begin{align}
\Delta W_{NN}\supseteq \frac{\tilde{c}_{\cal O} }{M_{\rm Pl}}\int\!d\theta^2\frac{\tilde{X}}{M_{\rm Pl}}\Bigl\{
  &\bar5_M\,10_F^{\,3}\,,\quad
  5_M^{\,2}\,\bar5_M\,\bar5_F\Bigr\},
 \nonumber\\[2pt]
\Delta K_{NN}\supseteq \frac{\tilde{c}_{\cal O} }{M_{\rm Pl}}\int\!d\theta^4\frac{\tilde{X}}{M_{\rm Pl}}\Bigl\{
  &\bar5_M^\dagger\,10_F^{\,2}\,,\quad
  5_M^\dagger\,\bar5_F\,10_F\Bigr\}
 +\text{h.c.},
 \label{eq:messenger-portal}
\end{align}
with proper coefficients $\tilde{c}_{\cal O}$.
Here the spurious chiral fields $X$ is a $U(1)_X$-neutral singlet, so the symmetry analysis
that follows is unchanged: the factor $\tilde{X}/M_{\rm Pl}$ only rescales the
coefficient of the dimension-five portal operators, which for $\langle \tilde{X}\rangle\neq0$ we write below as
$c_{\cal O}^{\rm eff}\equiv \tilde{c}_{\cal O}\langle \tilde{X}\rangle/M_{\rm Pl}$.
The first line contains superpotential
($F$-term) operators, while the second line contains Kahler ($D$-term)
operators. Each breaks the global $U(1)_m$ messenger number by exactly
one unit and is the leading Planck-suppressed interaction connecting the
messenger and visible sectors. 
The presence of these Planck-suppressed operators is an explicit model assumption, 
which is independent of the deflected AMSB setting and their coefficients could in principle be 
further suppressed by flavor symmetries or sequestering.

The operators of eq.~\eqref{eq:messenger-portal} fall into two classes,
which differ in which of these is realized. The two Kahler operators
are redundant in the effective-theory sense: each is removed by a shift
of one messenger chiral superfield, at the cost of inducing a trilinear
superpotential coupling (plus a harmless matter-only Kahler term of
order $M_{\rm Pl}^{-2}$). Removing $\bar5_M^\dagger10_F^2/M_{\rm Pl}$ by
$\bar5_M\to\bar5_M-M_{\rm Pl}^{-1}10_F^2$ turns the mass term
$M\bar5_M5_M$ into an induced Yukawa $\lambda\,5_M10_F^2$; removing
$5_M^\dagger\bar5_F10_F/M_{\rm Pl}$ by
$5_M\to5_M-M_{\rm Pl}^{-1}\bar5_F10_F$ induces
$\lambda\,\bar5_M\bar5_F10_F$. In both cases, the coupling $\lambda=M/M_{\rm Pl}$. The resulting
trilinear superpotential vertex with one messenger scalar and two matter
fermions admits no scalar-scalar two-body channel. So, in either case, the
scalar of the messenger superfield with the induced coupling decays into two
SM fermions with
$\Gamma\sim m_\varphi^3/(8\pi M_{\rm Pl}^2)$ and the involved LSP number for this decay is $B_{\rm LSP}=0$. The trilinear
couplings $\lambda\,5_M10_F^2$ and $\lambda\,\bar5_M\bar5_F10_F$
induced by these shifts carry the quantum numbers of a renormalizable
superpotential coupling, and $U(1)_X$ does not forbid them. The long
lifetime required by the late-decay scenario therefore rests on the
Planck suppression $\lambda=M/M_{\rm Pl}$, not on a symmetry. Note that an
unsuppressed coupling of this trilinear type would make the messenger decay
promptly, leaving $B_{\rm LSP}=0$ but removing the entropy-injection
mechanism. The absence of such a bare coupling is part of the model
assumption attached to eq.~\eqref{eq:messenger-portal}.

The superpotential operators $\bar5_M10_F^3$ and
$5_M^2\bar5_M\bar5_F$ involve four superfields and therefore mediate only
three-body decays. The superpartner component $\varphi_M\varphi_i\psi_j\psi_k$, with a single
MSSM superpartner and $R=-1$ for an $R$-even messenger scalar, is
forbidden, as are two-LSP final states. This class therefore also gives
$B_{\rm LSP}=0$, so the vanishing multiplicity is not a generic feature of
eq.~\eqref{eq:messenger-portal} but follows from the $R$-parity assignment.
In addition, the charges discussed below remove these operators except at
$a=5b$.

From the GUT uniform anomaly-free $U(1)_X$ charge assignments, a complete messenger pair carrying a single charge $q_M$ has
$q_X(\bar5_M)=-q_M$, so the two Kahler portals require
\begin{equation}
 \bar5_M^\dagger10_F^2:\ q_M=-2(a-b),\qquad
 5_M^\dagger\bar5_F10_F:\ q_M=2(a+b) .
 \label{eq:gut-uniform-portal}
\end{equation}
One of them therefore survives for $q_M=-2(a-b)$ or $q_M=2(a+b)$. The $U(1)_X$ charge for messengers should be non-trivial to provide non-vanishing gauge mediation contributions so as that $a\neq b$.
The superpotential portals require $q_M=3(a-b)$ and $q_M=-(a+3b)$, respectively. They are therefore absent except at $a=5b$, where the first superpotential portals opens for $q_M=2(a+b)$ and the second opens for $q_M=-2(a-b)$. 

For the scenario with GUT non-uniform anomaly-free $U(1)_X$ charge assignments for MSSM matter fields 
(that listed in Table.~\eqref{tab:fields}), the operators of eq.~\eqref{eq:messenger-portal} must be checked for $U(1)_X$ invariance term by term. Those terms that survive provide the
messenger-number-violating portal. Obviously, an invariant portal requires the charge choices
that differs between the triplet and doublet components of the messenger
${\bf5}$, which changes the messenger contribution to the Abelian beta
function.  Fortunately, invariant portal does exist for this scenario. Because $U(1)_X$ does not commute with $SU(5)$, the messenger
${\bf5}$ can carry different charges in its color-triplet and
weak-doublet components: $q_X(\Phi)=(q_3,q_2)$ for $5_M$ and $q_X(\bar{\Phi})=(-q_3,-q_2)$ for $\bar5_M$ by the vector-like mass term. The two Kahler operators of
eq.~\eqref{eq:messenger-portal} are therefore invariant for
\begin{align}
 \bar5_M^\dagger10_F^2 &: \quad
 q_X(\Phi)=\bigl(1/9,-1/2\bigr)\ \text{or}\ \bigl(-5/9,-1/2\bigr),
 \label{eq:q-C}\\
 5_M^\dagger\bar5_F10_F &: \quad
 q_X(\Phi)=\bigl(5/9,\,1/2\bigr).
 \label{eq:q-D}
\end{align}
The two possible assignments of $q_3$ in \eqref{eq:q-D} correspond to different contractions of $10_F10_F$. 
The assignment \eqref{eq:q-D} is economical, which states that $5_M$ carries the same $U(1)_X$ charges as the Standard-Model
$\bar5_F=\{d^c,L\}$. The two superpotential operators are not
invariant with it. So the $R$-even Kahler portal is the only one present and
$B_{\rm LSP}=0$. We have three remarks: First, no uniform assignment is
possible in this scenario, because the charges allowed for the triplet component
($1/9$, $5/9$, $-5/9$) and for the doublet component ($\pm1/2$) are
never equal. Second, no assignment satisfies both Kahler operators. One allowed
portal is sufficient, since a single messenger-number-violating coupling
already makes the messenger decay. Third, the price is a change in the
messenger contribution to the Abelian beta function,
\begin{equation}
 \Delta b_X=2N\bigl(3q_3^2+2q_2^2\bigr)
 =
 \begin{cases}
  2.85\,N, & q_X(\Phi)=\bigl(5/9,1/2\bigr)\ \text{or}\
            \bigl(-5/9,-1/2\bigr),\\
  1.07\,N, & q_X(\Phi)=\bigl(1/9,-1/2\bigr),
 \end{cases}
 \label{eq:DbX-split}
\end{equation}
against $\Delta b_X=0.4N$ for the uniform assignment with $q_M=0.2$. The two $F$-term
operators remain forbidden for these charges, which is what makes
$B_{\rm LSP}=0$ a statement about the continuous symmetry rather than
only about parity. First, the operator $\bar5_M10_F^3$ requires the ${\bf5}$ of
$10_F^3$ to carry $-q_X(\bar5_M)$. However, in the triplet sector, the values
available from the matter charges are $4/9$ and $10/9$, whereas the
charge demanded by the economical assignment \eqref{eq:q-D} is $5/9$.
Second, the operator $5_M^2\bar5_M\bar5_F$ needs
$q_X(5_M)=-q_X(\bar5_F)$ component by component. But $5_M$
carries exactly the charges of the $\bar5_F=\{d^c,L\}$.

We will focus mainly on the scenarios with GUT uniform anomaly-free $U(1)_X$ charge assignments in our following discussions.

After expanding in component fields,
the leading decay of a messenger scalar $\varphi$ into MSSM
states via the messenger portal can be parameterized as
\begin{equation}
 \Gamma_{\rm portal}\sim\frac{|c_{\cal O}^{\rm eff}|^2}{8\pi}
 \frac{m_\varphi^3}{M_{\rm Pl}^2},\quad{\rm with}~~ c_{\cal O}^{\rm eff}\equiv \tilde{c}_{\cal O}\frac{\langle \tilde{X}\rangle}{M_{\rm Pl}}.
\label{eq:messenger-portal-decay}
\end{equation}
Every value of $|{c}_{\cal O}|$ quoted in this paper refers to $|c_{\cal O}^{\rm eff}|$.
For $m_\varphi=10^{10}\GeV$ with $|c_{\cal O}^{\rm eff}|\simeq3\times10^{-8}$,
its lifetime is $0.1 \,\mathrm{s}$, precisely the range relevant to Big Bang Nucleosynthesis (BBN), 
as well as for late entropy injection and non-thermal Higgsino production to be discussed in section~\ref{sec:relic}.

The messenger fermion can decay into the messenger scalar plus a gaugino.
This is a tree-level gauge vertex, so the width carries no
supersymmetry-breaking suppression; it is set by the mass splitting inside
the messenger multiplet,
\begin{equation}
 \Gamma(\chi_\phi\to\phi\,\lambda_a)
 \sim\sum_a\frac{g_a^2C_a}{16\pi}\,M_{\rm mess}
 \left[\frac{(1+d)F_\phi}{M_{\rm mess}}\right]^{2},
\label{eq:messenger-gaugino-width}
\end{equation}
where the bracket is the two-body phase space,
$1-m_\phi^2/M_{\rm mess}^2=(1+d)F_\phi/M_{\rm mess}$.
The messenger fermion can also decay into a messenger scalar plus a goldstino through the supersymmetry-breaking spurion coupling. With a
spurion of mass $M_{\rm mess}$ and messenger--spurion coupling $\lambda$,
this goldstino channel has width~\cite{Dalianis}
\begin{equation}
 \Gamma_\lambda(\chi_\phi\to\phi\,\psi_G)
 =\frac{\lambda^4k^4F^2}{4\pi M_{\rm mess}^3},
\label{eq:messenger-goldstino-width}
\end{equation}
with $k=F_{\tilde{X}}/F$ the fraction of supersymmetry breaking felt by the messenger particles. In our simple deflected AMSB scenario, 
the F-term for pseudo-moduli driven by the deflection is given by $F_{\tilde{X}}/\tilde{X}=(d+1)F_\phi$, so that 
\beqa k=\f{F_{\tilde{X}}}{F}=\f{F_{\tilde{X}}}{m_{3/2}M_{Pl}}\sim\f{(d+1)\tilde{X}}{M_{Pl}} \ll 1~. \eeqa
With the Planck mass $M_{\rm Pl}=1.22\times10^{19}\GeV$ used throughout,
$k\simeq8.2\times10^{-9}$ for $d=3$, $\lambda=0.4$ and
$M_{\rm mess}=10^{10}\GeV$.

Therefore, the decay channel of messenger fermion  into  gaugino and scalar messenger is always the dominant one.

One feature of the deflected-AMSB realization must be stated explicitly. The gravitino mass is set by the same order parameter as
the soft terms but carries no loop suppression,
$m_{3/2}=F_\phi$, whereas the gaugino and scalar masses are
one-loop, $M_i\propto F_\phi g_i^2(b_i-Nd)/16\pi^2$. The ratio is
therefore $m_{3/2}/m_{\rm soft}\sim16\pi^2/[g^2(b-Nd)]$, of order
$10^2$--$10^3$ for general choices of parameters. The gravitino is consequently heavier
than the visible superpartners, contrary to minimal gauge-mediation scenarios, where the gravitino is the
lightest $R$-odd state and the stable dark matter candidate. In the
deflected-AMSB framework used here, the gravitino is unstable and decays
gravitationally into a superpartner plus a Standard-Model particle,
\begin{equation}
 \Gamma_{3/2}=\frac{12m_{3/2}^3}{32\pi M_{\rm Pl}^2},
\label{eq:g3-2-width}
\end{equation}
on a timescale much longer than the messenger decay widths of
eqs.~\eqref{eq:messenger-portal-decay}.
The messenger sector plays multiple cosmological
roles: its late decays can produce Higgsinos non-thermally and
simultaneously inject entropy that dilutes any pre-existing
abundance. 

The only visible-sector route to the lightest Higgsino proceeds
through the dimension-five portal decay of
eq.~\eqref{eq:messenger-portal-decay}: the messenger decays into MSSM
states through the messenger-number-violating portal, and the subsequent
cascade terminates in the Higgsino LSP. The portal width is suppressed by
$(m_\varphi/M_{Pl})^2$ relative to a gauge-strength interaction, but it
is the only available visible-sector decay channel in the absence of
messenger--matter Yukawa mixing, and it is this suppression that makes the
scalar messenger long lived. The competing goldstino channel of
eq.~\eqref{eq:messenger-goldstino-width} carries four powers of $k$ and is
negligible, as shown in section~\ref{sec:gravitino}, so no
two-step chain through the gravitino contributes to the yield. In
deflected AMSB the gravitino is heavier than the visible superpartners and
unstable, unlike minimal gauge mediation where it is the LSP and stable;
its abundance here is the thermal one of
eq.~\eqref{eq:tr-criterion}, three orders of magnitude below the thermal
Higgsino yield, and it is diluted in the same way. 

When a component does dominate before decaying, its entropy release is
conventionally written as the ratio of the would-be domination
temperature to the decay temperature
\begin{equation}
 \Delta_{\rm mess}=\frac{T_{\rm dom}}{T_{\rm decay}}.
\label{eq:dmess}
\end{equation}
Here, the would-be domination temperature $T_{\rm dom}$ is defined to be the value when 
the energy density $\rho_\phi=M Y s$ for a non-relativistic component of mass $M$ and dark matter comoving number density (yield) $Y$  
equal to the radiation 
\begin{equation}
 \rho_{\rm rad} =\frac{\pi^2g_*T^4/30}{2\pi^2g_{*s}T^3/45}{s}
 =\frac34\frac{g_*}{g_{*s}}T {s}.
 \label{eq:rhorads}
\end{equation}
Equating the two defines 
\begin{equation}
 T_{\rm dom}=\frac43\frac{g_{*s}}{g_*}\,MY.
 \label{eq:tdom}
\end{equation}
 A component that survives below $T_{\rm dom}$
dominates temporarily, and its decay then releases entropy that dilutes
every earlier abundance by $\Delta_d\simeq T_{\rm dom}/T_{\rm decay}$,
whereas no dilution occurs if
$T_{\rm decay}>T_{\rm dom}$\footnote{Note that eq.~\eqref{eq:dmess} is a ratio of temperatures rather than
the entropy ratio itself. Writing $\epsilon\equiv\rho_X/\rho_{\rm rad}$ just before
the decay, one finds $\Delta_{\rm mess}=\epsilon(1+\epsilon)^{-1/4}$. For the
true dilution factor of eq.~\eqref{eq:delta-def}, $\Delta_d=(1+\epsilon)^{3/4}$,
so that $\Delta_d/\Delta_{\rm mess}=1+1/\epsilon$. The two agree only in the
strongly dominated regime $\epsilon\gg1$, while $\Delta_{\rm mess}<1$ simply
signals the absence of dilution ($\Delta_d\simeq1$) rather than a suppression.}.

Applying to the late decaying messenger, we have $T_{\rm dom}\simeq \tfrac43M_{\rm mess}Y_{\phi_{LM}}$, which holds for $g_*=g_{*s}$. 
 For $M_{\rm mess}=10^{10}\GeV$ and
$Y_{\phi_{LM}}=8.3\times10^{-9}$ (the benchmark of
eq.~\eqref{eq:heavy-benchmark}), eq.~\eqref{eq:tdom} gives $T_{\rm dom}\simeq110\GeV$. The fermionic messenger contributes no gravitino yield (section~\ref{sec:gravitino}), 
yielding $\Delta_d=Y_{\rm th}/Y_{\rm req}\simeq6.6\times10^{3}$ for $B_{\rm LSP}=0$, independent of $m_X$. 
Note that $\Delta_{\rm mess}$ is a function of the
messenger scale and couplings rather than a free parameter. In contrast to gauge-mediated scenarios where the gravitino
is the dark matter~\cite{Dalianis} or the lightest messenger is
stable~\cite{Dimopoulos:1996nbb}, the present framework uses the
messenger as an unstable source, whose decay simultaneously injects
Higgsinos and entropy. The two effects compete in the same final-yield
equation.

\section{Messenger decays and the Higgsino relic abundance}
\label{sec:relic}
In the deflected-AMSB framework, the dark matter is the Higgsino. 
Late messenger decays serve two simultaneous purposes: they
inject Higgsinos that supplement or replace the thermal relic, and they
release entropy that can dilute an overabundant thermal population.

We use three reference Higgsino masses to distinguish the required
cosmological histories. The thermal abundances in table~\ref{tab:thermal}
are the benchmark estimates underlying the present analysis, calibrated to
$\Omega_\chi h^2=0.1200$ at $m_\chi=1.1\TeV$. They are not the output of
a completed scan of the singlet-extended $U(1)_X$ neutralino sector. In particular, applying
them requires a Higgsino LSP with sufficiently decoupled extra states.
The reference numerical treatment includes an effective Sommerfeld treatment and
assigns a factor-of-two normalization uncertainty to the thermal
abundance.

\begin{table}[tbp]
\centering\small
\begin{tabular}{lccccc}
\toprule
Regime & $m_\chi$ & $\Omega_{\rm th}h^2$ & $Y_{\rm th}$ & $Y_{\rm req}$ & $Y_{\rm th}/Y_{\rm req}$\\
\midrule
I & $500\GeV$ & $0.0260$ & $1.897\!\times\!10^{-13}$ & $8.744\!\times\!10^{-13}$ & $0.217$\\
II & $1.1\TeV$ & $0.1200$ & $3.975\!\times\!10^{-13}$ & $3.975\!\times\!10^{-13}$ & $1.000$\\
III & $100\TeV$ & $792.5$ & $2.888\!\times\!10^{-11}$ & $4.372\!\times\!10^{-15}$ & $\simeq6.6\!\times\!10^3$\\
\bottomrule
\end{tabular}
\caption{Reference thermal abundances and the yields required for
$\Omega_{\rm DM}h^2=0.12$. }
\label{tab:thermal}
\end{table}

At $500\GeV$, the thermal component supplies approximately $22\%$ of the
observed abundance, so a single-component Higgsino scenario requires additional
production. Near $1.1\TeV$, no late modification is necessary. Additional
production must compensate any entropy dilution. At $100\TeV$, the
thermal dark matter abundance instead requires a dilution of about
$6.6\times10^3$ in the absence of subsequent LSP production. These are
three parameter regimes, not one common parameter point.

Let $X$ denote the physical late-decaying messenger state and
$m_X\simeq M_{\rm mess}$. We write $B_{\rm LSP}$ for the mean number of stable Higgsinos produced
per decay, including cascades. Note that it is a multiplicity, not necessarily
a branching fraction bounded by one. The decay is assumed to occur
after Higgsino chemical freeze-out.

\subsubsection*{Instantaneous-decay kinematics}

The following quantities characterise the transition. Let $T_b$ be the
radiation-bath temperature immediately before the decay, and let
$T_d$ be the post-decay reheating temperature, defined as the
temperature the radiation bath would attain if the entire energy density
of $X$ were converted instantaneously into relativistic degrees of
freedom. Define the energy-density ratio just before decay as
\begin{equation}
 \epsilon\equiv\frac{\rho_X}{\rho_{\rm rad}}\bigg|_{T=T_b}.
 \label{eq:epsilon-def}
\end{equation}
The comoving yield of $X$ before decay is
$Y_X^{\rm before}\equiv n_X/s(T_b)$, where $s(T)$ is the entropy density.
Throughout this subsection we assume that the effective relativistic
degrees of freedom for energy density ($g_{*\rho}$) and entropy density
($g_{*s}$) are equal, and do not change across the transition. This is
valid for $T\sim\mathcal{O}(\text{MeV})$, where
$g_{*\rho}\simeq g_{*s}\simeq10.75$. Near particle-mass thresholds
where $g_{*\rho}\neq g_{*s}$, the formulae below acquire correction
factors of order unity.

Instantaneous conversion of $\rho_X$ into radiation gives
$$\rho_{\rm rad}(T_d)=\rho_{\rm rad}(T_b)+\rho_X
=\rho_{\rm rad}(T_b)(1+\epsilon).$$ Since $\rho_{\rm rad}\propto g_*T^4$
with fixed $g_*$,
\begin{equation}
 T_d^4=T_b^4(1+\epsilon)\quad\Rightarrow\quad
 T_b=T_d(1+\epsilon)^{-1/4}.
 \label{eq:Td-Tb}
\end{equation}

The entropy density scales as $s\propto g_*T^3$. The ratio of the
post-decay to pre-decay entropy density defines the dilution factor:
\begin{equation}
 \Delta_d\equiv\frac{s(T_d)}{s(T_b)}
 =\left(\frac{T_d}{T_b}\right)^3
 =(1+\epsilon)^{3/4}.
 \label{eq:delta-def}
\end{equation}
Any pre-existing relic number density normalized to entropy is suppressed by
$1/\Delta_d$ after the transition: $Y^{\rm after}=Y^{\rm before}/\Delta_d$.

For non-relativistic $X$, the energy density
$\rho_X=m_Xn_X=m_XY_X^{\rm before}s(T_b)$.
Using $\rho_{\rm rad}(T_b)/s(T_b)=3T_b/4$, which follows from
$\rho_{\rm rad}=\pi^2g_{*\rho}T^4/30$ and $s=2\pi^2g_{*s}T^3/45$
when $g_{*\rho}=g_{*s}$,\footnote{Strictly, the energy-density and
entropy-density effective degrees of freedom differ near particle-mass
thresholds. At $T_d=4\,\MeV$ all relevant species (photons, neutrinos,
$e^\pm$) are fully relativistic and
$g_{*\rho}\simeq g_{*s}\simeq10.75$, so the equality holds to high
accuracy. For $T_d$ near the electron-annihilation temperature
($\sim0.5\,\MeV$) or below, one should replace $3T/4$ by
$(3T/4)(g_{*\rho}/g_{*s})$ and track both functions separately.} the
energy-density ratio becomes
\begin{equation}
 u\equiv m_XY_X^{\rm before}
 =\frac{\rho_X}{s(T_b)}
 =\epsilon\,\frac{\rho_{\rm rad}(T_b)}{s(T_b)}
 =\frac34\epsilon T_b.
 \label{eq:u-def}
\end{equation}
Substituting $T_b$ from eq.~\eqref{eq:Td-Tb}:
\begin{equation}
 u=\frac34\epsilon\,T_d(1+\epsilon)^{-1/4}.
 \label{eq:u-Td}
\end{equation}
In the domination limit $\epsilon\gg1$, this simplifies to
$u\simeq3 T_d \Delta_d/4$.

\subsubsection*{Final-yield equation}

In the instantaneous-decay approximation the total Higgsino yield after
the transition is the sum of the diluted thermal component and the
injected non-thermal component:
\begin{align}
 \Omega_\chi h^2&=2.7445\times10^8
       \left(\frac{m_\chi}{\mathrm{GeV}}\right)Y_\chi,
 &Y_\chi&=\frac{Y_{\rm th}+B_{\rm LSP}Y_X^{\rm before}}{\Delta_d}.
 \label{eq:yield}
\end{align}
The multiplicity $B_{\rm LSP}$ is inclusive over the full cascade and is
not bounded above by unity. A specific decay may produce
one LSP, but that is not a general property of the framework. 
This expression does not by itself include
post-injection re-annihilation. A quantitative calculation in a regime
where that process matters must solve the decay and annihilation terms
together.

The injected yield after dilution can be rewritten entirely in terms of
$T_d$, $m_X$ and $\epsilon$ by combining eqs.~\eqref{eq:delta-def}
and~\eqref{eq:u-Td}:
\begin{equation}
 Y_{\rm NT}=\frac{B_{\rm LSP}Y_X^{\rm before}}{\Delta_d}
 =\frac34 B_{\rm LSP}\frac{T_d}{m_X}\frac{\epsilon}{1+\epsilon}.
 \label{eq:nonthermal-yield}
\end{equation}
When the $X$ source dominates ($\epsilon\gg1$), $Y_{\rm NT}$ approaches $3B_{\rm LSP}T_d/(4m_X)$ and no longer decreases when the source
abundance is increased to generate more entropy. Thus production and
dilution cannot be treated as independent adjustments of the same
dominating source. Benchmark inputs and numerical cross-checks are
collected in appendix~\ref{app:dilution}.

Subtracting the initial thermal yield makes the net effect explicit:
\begin{equation}
 Y_\chi-Y_{\rm th}=
 \frac{B_{\rm LSP}Y_X^{\rm before}-(\Delta_d-1)Y_{\rm th}}{\Delta_d}.
 \label{eq:net-messenger-effect}
\end{equation}
A messenger decay is net productive or net diluting according to the
sign of the numerator. Its lifetime controls when the decay occurs,
while its abundance and decay channels determine the competition
between the two terms. If it decays before freeze-out, or if the
injected population subsequently annihilates efficiently, the simple
post-freeze-out formula must be replaced by the corresponding thermal
or continuous-decay evolution.

\subsection{Light Higgsinos: non-thermal contribution}
\label{sec:light-production}

For a $500\GeV$ Higgsino the thermal yield falls short of the yield required by the full abundance,
$Y_{\rm req}$. Obtaining the full dark-matter abundance requires
\begin{equation}
 B_{\rm LSP}Y_X^{\rm before}=\Delta_d  Y_{\rm req}-Y_{\rm th}.
 \label{eq:production-target}
\end{equation}
With negligible dilution, the missing yield is approximately
$6.85\times10^{-13}$. The source need not dominate the energy
density, and its direct or cascade decays into MSSM states supply the
missing Higgsinos through the portal cascade of
eq.~\eqref{eq:messenger-portal-decay}. In this regime the messenger acts
as a non-thermal Higgsino source rather than an entropy dilution source. The
portal-mediated decay is the minimal mechanism that breaks messenger
number and allows the messenger to feed the Higgsino abundance. For
appreciable entropy release, the required injected supplementation increases
according to eq.~\eqref{eq:production-target}, which indicates that any entropy injected by
the same messenger must be compensated by additional Higgsino production.
The messenger abundance and LSP multiplicity must be computed from its
interactions, while the target supplementation is not itself a decay-model prediction.

\subsection{$1.1\TeV$ Higgsinos: balanced abundance}
\label{sec:thermal-preservation}

Near $1.1\TeV$, the thermal abundance already matches the
observed dark matter relic density. A negligible late source leaves this result intact.
If messenger decays do modify the entropy, production must compensate
for the lost thermal yield:
\begin{equation}
 B_{\rm LSP}Y_X^{\rm before}=Y_{\rm req}(\Delta_d-1).
 \label{eq:balanced-source}
\end{equation}
This is a balance condition, not a generic consequence of $1.1\TeV$ Higgsino mass. Arbitrary additional production would instead
overclose the universe.

\subsection{Heavy Higgsinos: entropy dilution and regeneration}
\label{sec:heavy-dilution}

At $m_\chi=100\TeV$, the thermal production of Higgsino dark matter exceeds $Y_{\rm req}$
by approximately $6.6\times10^3$, so an entropy dilution
$\Delta_d\gtrsim6.6\times10^3$ is required. This factor cannot be replaced
by post-injection annihilation: annihilation can at best drive the
comoving abundance back to its freeze-out value, which is itself
over-abundant by the same factor. A genuine
entropy source is therefore unavoidable. A messenger that dominates the
energy density before decay can supply it, but its own decays into
Higgsinos, through the portal cascade of
eq.~\eqref{eq:messenger-portal-decay}, are constrained
independently. Note that the same messenger that dilutes the thermal Higgsino abundance
may also regenerate it.

The choice of $T_d$ is constrained from below by BBN.  Hadronic and electromagnetic decay products injected during or
after BBN ($T\sim0.1$--$1\,\MeV$, $t\sim1$--$10^3\,\mathrm{s}$) photodissociate
deuterium, overproduce ${}^6\mathrm{Li}$, and alter the neutron-to-proton
ratio, conflicting with observed light-element abundances.  Detailed
calculations~\cite{KKM2005,Jedamzik2006} show that for order-one hadronic
branching ratios, the decay must complete before $T\sim4$--$5\,\MeV$
($\tau\lesssim0.1\,\mathrm{s}$) to avoid these bounds.  A higher $T_d$
is always BBN-safe but yields a smaller dilution factor $\Delta_d$, because
the source dominates for a shorter interval and $\epsilon$ is reduced. The benchmark $T_d=4\,\MeV$ therefore represents the \emph{maximum
dilution} consistent with BBN safety for generic decay channels. It is
not a prediction of the dAMSB framework, but a phenomenological favored
point near the BBN lower bound.  A complete treatment must verify the
BBN constraints for the specific branching ratios of the portal operator
eq.~\eqref{eq:messenger-portal}. At $T_d=4\,\MeV$, the weak interaction rate satisfies $\Gamma_\nu/H\approx20$, 
so neutrinos remain in equilibrium with the plasma and the effective number of relativistic species stays at its SM value. 

The post-decay temperature is related to the total decay width by the
Friedmann equation evaluated at $\Gamma=H$:
\begin{equation}
 T_d=\left(\frac{90}{\pi^2 g_*}\right)^{\!1/4}\!\sqrt{\Gamma M_{\rm Pl}}
 \;\simeq\;1.74\,g_*^{-1/4}\sqrt{\Gamma M_{\rm Pl}},
 \label{eq:Td-Gamma}
\end{equation}
where the numerical prefactor uses $g_*=10.75$.  For $T_d=4\,\MeV$, this
gives $\Gamma\simeq7.1\times10^{-24}\,\GeV$ and
$\tau\simeq9.3\times10^{-2}\,\mathrm{s}$. With this benchmark, for
$\Delta_d\simeq2.8\times10^4$, we have $u\simeq3 T_d \Delta_d/4=82.8\GeV$, which further gives
$\epsilon\simeq8.3\times10^5$ in eq.~\eqref{eq:delta-def}. An alternative pressure-based integration of the same entropy
relation gives $\Delta_d=2.94\times10^4$ and we use the analytic value
$2.8\times10^4$ throughout. The
freeze-out input must also be recomputed if messenger domination already
changes the expansion rate during chemical decoupling.

The decay width and the source abundance are independent benchmark
inputs. A total width $\Gamma=7.1\times10^{-24}\GeV$, for instance two
partial widths of $3.567\times10^{-24}\GeV$, gives
$\tau=9.2\times10^{-2}\,\mathrm{s}$ and, through
eq.~\eqref{eq:Td-Gamma}, $T_d\simeq4\MeV$. The source abundance will be fixed by $u\simeq3 T_d \Delta_d/4$ to be $u\equiv m_XY_X^{\rm before}=82.8\GeV$. Once the source mass is chosen, we have
\begin{equation}
 Y_X^{\rm before}=\frac{u}{m_X}
 =8.28\times10^{-5}\left(\frac{10^6\GeV}{m_X}\right).
 \label{eq:required-YX}
\end{equation}
A source of $m_X=10^{10}\GeV$ therefore requires
$Y_X^{\rm before}=8.3\times10^{-9}$, the value used in
appendix~\ref{app:dilution}. A lighter source needs a proportionally
larger abundance. The corresponding equivalent coupling in the normalization of
eq.~\eqref{eq:messenger-portal-decay} is $|c_{\cal O}^{\rm eff}|\simeq3\times10^{-8}$ for
$m_X=10^{10}\GeV$.

The non-thermal component must satisfy
\begin{equation}
 \frac{B_{\rm LSP}Y_X^{\rm before}}{\Delta_d}\leq Y_{\rm req},
 \qquad
 Y_{\rm req}=\frac{0.12}{2.7445\times10^8(m_\chi/\mathrm{GeV})}.
 \label{eq:yieldcriterion}
\end{equation}
In the dominating-source limit, a positive surviving thermal component
requires
\begin{equation}
 B_{\rm LSP}< B_{\max}\equiv\frac43\frac{m_XY_{\rm req}}{T_d}.
 \label{eq:bmax}
\end{equation}
At $m_\chi=100\TeV$, $m_X=10^{10}\GeV$, $T_d=4\MeV$,
and $Y_{\rm req}=4.372\times10^{-15}$ lead to $B_{\max}=1.46\times10^{-2}$.
An order-one LSP multiplicity would instead give
$Y_{\rm NT}=3.0\times10^{-13}$, seventy times the abundance required by eq.~\eqref{eq:yieldcriterion}.
Since $B_{\max}\propto m_X$, a heavier source relaxes the bound: the cascade
multiplicity $B_{\rm LSP}\simeq10^{-2}$ needs only $m_X\gtrsim7\times10^{9}\GeV$. Requiring
$B_{\rm LSP}=1$ instead needs $m_X\gtrsim6.9\times10^{11}\GeV$, where
$T_d=4\MeV$ corresponds to $|c_{\cal O}^{\rm eff}|\simeq6\times10^{-11}$,
i.e.\ $\langle \tilde{X}\rangle\simeq7\times10^{8}\GeV$ for $c_{\cal O}$ of order unity.

Since $B_{\rm LSP}$ counts the stable Higgsinos produced per
decay, the indirect decay into Higgsino via SM cascade decay with branching fraction $\mathrm{Br}$
gives $B_{\rm LSP}=2\mathrm{Br}$, and eq.~\eqref{eq:bmax} reads
$\mathrm{Br}<7.3\times10^{-3}$ at the $m_X=10^{10}\GeV$ benchmark. The
decay vertex itself carries no Higgsino. The portal into SM particles injects energy $m_X$, and the following cascade decay
pair-produces superpartners through gauge interactions, the gauginos among them then
decaying into Higgsinos. Since the superpartners are $R$-odd, they are produced only in
pairs, so the
multiplicity is of order $(\alpha_s/4\pi)^2\ln^2[m_X/(2m_{\tilde{g}})]$ times the number of partons:
$B_{\rm LSP}\simeq(0.3$--$1.3)\times10^{-2}$ for $m_X=10^{7}$--$10^{10}\GeV$, the
threshold factor $\beta^3$ of the pair emission making this an upper bound. No
symmetry of the messenger sector forbids it. The almost vanishing multiplicity
is therefore a requirement on the source rather than a consequence of
the messenger sector. The vanishing multiplicity is met by construction when the source lies
below the pair-production threshold, $m_{\rm source}<2m_\chi$, as for the
singlet scalar of appendix~\ref{sec:sigma}.

 For
$B_{\rm LSP}=10^{-2}$, eq.~\eqref{eq:bmax} requires $m_X\gtrsim7\times10^{9}\,\GeV$
at $T_d=4\,\MeV$, so we take $m_X=10^{10}\,\GeV$. Solving eq.~\eqref{eq:yield} for
$\Omega_\chi h^2=0.12$ gives
\begin{equation}
 Y_X^{\rm before}=\frac{Y_{\rm th}+Y_{3/2}^{\rm th}-Y_{\rm req}}{B_{\rm max}-B_{\rm LSP}}
 \simeq8.3\times10^{-9},\quad u\simeq83\,\GeV,\quad
 \Delta_d\simeq2.8\times10^{4},
\label{eq:heavy-benchmark}
\end{equation}
where $B_{\rm LSP}<B_{\rm max}=1.46\times10^{-2}$, and the
required portal coupling is $|c_{\cal O}^{\rm eff}|\simeq3\times10^{-8}$, i.e.\
$\langle\tilde{X}\rangle\simeq4\times10^{11}\,\GeV$ for $c_{\cal O}$ of order unity. 
First, this solution is specific to the overabundant
branch. Second, what the BBN discussion above constrains is the
energy injected per photon, which is set by $u$ and not by $m_X$: with
$u\simeq83\,\GeV$ the bounds of
refs.~\cite{KKM2005,Jedamzik2006} are met for the same $T_d$, and only the
numerical
coefficient of $\Delta_d$ changes.

\begin{figure}[tbp]
 \centering
 \includegraphics[width=\textwidth]{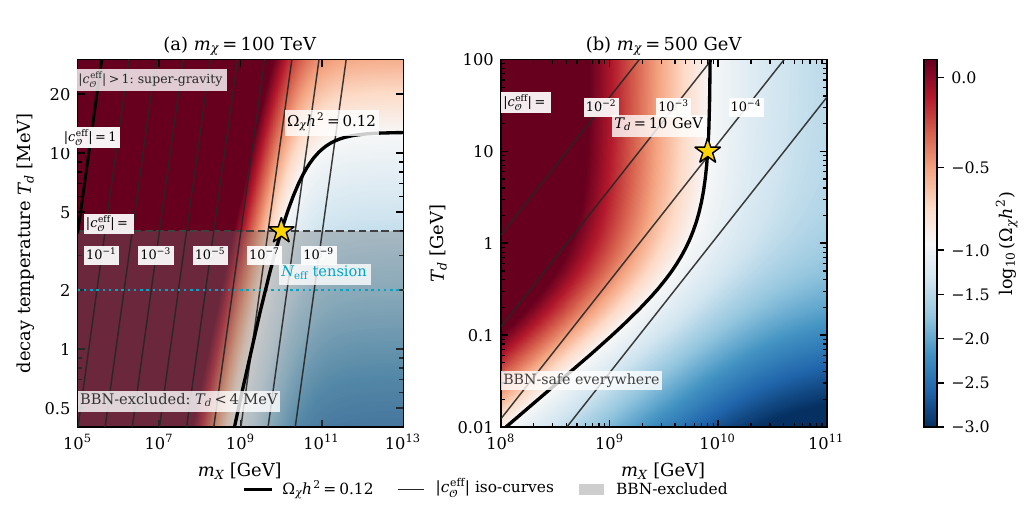}
 \caption{Decay temperature required by the relic density in the
$(m_X,\,T_d)$ plane, in two panels. Panel (a): $m_\chi=100\TeV$ with
$B_{\rm LSP}=10^{-2}$ and $u=m_XY_X^{\rm before}=82.8\GeV$, including the
thermal gravitino yield of eq.~\eqref{eq:tr-criterion}; the vertical scale is
in MeV. Panel (b): the light branch, $m_\chi=500\GeV$ with
$B_{\rm LSP}=1.3\times10^{-2}$ and $u=0.45\GeV$; the vertical scale is in GeV,
and the whole plotted range is BBN-safe. Colour gives
$\log_{10}(\Omega_\chi h^2)$ on a scale centred on the observed value, the
thick line is $\Omega_\chi h^2=0.12$, and the stars mark the two working
points. The thin curves show the effective portal coupling the decay then
requires, $|c_{\cal O}^{\rm eff}|=T_d\sqrt{8\pi M_{\rm Pl}/(0.9231\,m_X^3)}$;
at the $100\TeV$ star $|c_{\cal O}^{\rm eff}|\simeq3\times10^{-8}$, and at the
$500\GeV$ star $|c_{\cal O}^{\rm eff}|\simeq1\times10^{-4}$. In
panel (a) the grey band below the BBN bound, $T_d\simeq4\MeV$, is excluded by
light-element photodissociation for order-one hadronic branching ratios, and
the cyan dotted line marks the Planck bound on $N_{\rm eff}$.}
 \label{fig:parameter-space}
\end{figure}

The two terms of eq.~\eqref{eq:yield} are not independent, because
$\Delta_d$ and $Y_X^{\rm before}$ are fixed by the same source. In the
domination limit $\Delta_d\simeq4m_XY_X^{\rm before}/(3T_d)$, we have
\begin{equation}
 Y_\chi\simeq\frac{3T_d}{4m_X}
 \left(\frac{Y_{\rm th}}{Y_X^{\rm before}}+B_{\rm LSP}\right)
 \;\longrightarrow\;\frac{3T_d}{4m_X}\,B_{\rm LSP}~,
 \quad\text{as }Y_X^{\rm before}\to\infty .
 \label{eq:injection-floor}
\end{equation}
Increasing the source abundance raises the entropy release and the
injected LSP number density in the same proportion, and the two effects cancel
exactly. What survives is a floor set by $T_d/m_X$, equal to
$3.0\times10^{-15}$ for the benchmark values $B_{\rm LSP}=10^{-2}$ and
$m_X=10^{10}\GeV$. The decay temperature that the two branches require is displayed in fig.~\ref{fig:parameter-space}.

The injected Higgsino population is far from chemical equilibrium at $T_d$. For
$B_{\rm LSP}=10^{-2}$, the post-dilution yield of
eq.~\eqref{eq:nonthermal-yield} is $Y_{\rm NT}=3.0\times10^{-15}$, so that
$n_\chi=Y_{\rm NT}s(T_d)\simeq9.1\times10^{-22}\GeV^3$. Taking into account the Coulomb Sommerfeld factor, we can estimate $\langle\sigma v\rangle\simeq1.8\times10^{-10}\GeV^{-2}$ and hence $\Gamma_{\rm ann}=n_\chi\langle\sigma v\rangle\simeq1.6\times10^{-31}\GeV$ against
$H(T_d)=7.1\times10^{-24}\GeV$. The ratio
$\Gamma_{\rm ann}/H\simeq2.3\times10^{-8}$ decreases as the universe
cools, so the injected population does not annihilate within a Hubble time. The appropriate bound is not the freeze-out yield, but the
annihilation-limited yield at $T_d$,
\begin{equation}
 Y_{\rm floor}=\frac{H}{s\langle\sigma v\rangle}\bigg|_{T_d}
 \simeq\frac{7.1\times10^{-24}}
             {3.0\times10^{-7}\times1.8\times10^{-10}}
 \simeq1.3\times10^{-7}.
 \label{eq:Yfloor}
\end{equation}
This equation is actually a rate-matching criterion for the maximum depletion of abundance via annihilation, 
that is, how low the abundance can be maximally suppressed by annihilation (rather than the Boltzmann freeze-out solution).
The floor separates two regimes. For $Y>Y_{\rm floor}$,
 the annihilation rate $\Gamma_{\rm ann}=Y s\langle\sigma v\rangle$ exceeds the expansion
rate $H$, so the abundance is driven down towards $Y_{\rm floor}$; for
$Y<Y_{\rm floor}$, it falls short of $H$, annihilations freeze out, and
$Y$ no longer decreases. $Y_{\rm floor}$ is therefore the lowest yield
that annihilation can reach at $T_d$. For a $Y$ value below it, the abundance can only be
reduced by entropy dilution, not by further annihilation. The rate $Y_{\rm floor}\simeq1.3\times10^{-7}$ is the lowest abundance that dark matter annihilation can reach at this temperature. Because $H/s\propto1/T$, this floor exceeds the rate $Y_{\rm th}=2.9\times10^{-11}$ by a factor $4.5\times10^3$ and $Y_{\rm req}$ by $3\times10^7$. 

The same equations describe the opposite situation, an underabundant thermal
population, which is the case of the light Higgsino of section~\ref{sec:light-production}. There
$Y_{\rm th}<Y_{\rm req}$, so eq.~\eqref{eq:yield} can be satisfied only if the source injects faster
than it dilutes, that is, when
\begin{equation}
 B_{\rm LSP}>B_0\equiv\frac{4m_XY_{\rm req}}{3T_d},
 \label{eq:bmin}
\end{equation}
which requires an early decay. For $m_\chi=500\GeV$, the deficit is a factor of $4.6$
($Y_{\rm th}=1.9\times10^{-13}$ against $Y_{\rm req}=8.7\times10^{-13}$). So, with
$B_{\rm LSP}\simeq1.3\times10^{-2}$ and $m_X=8\times10^{9}\GeV$, eq.~\eqref{eq:bmin} is met for
$T_d\gtrsim0.7\GeV$. A representative point is $T_d=10\GeV$, where
$Y_X^{\rm before}\simeq5.8\times10^{-11}$, $u\simeq0.46\GeV$, $\epsilon\simeq6\times10^{-2}$ and
$\Delta_d\simeq1.05$: the dilution is negligible, the injected yield
$Y_{\rm NT}\simeq6.9\times10^{-13}$ supplies the whole deficit, and $\Omega_\chi h^2=0.12$ follows.
The corresponding coupling is $|c_{\cal O}^{\rm eff}|\simeq1.1\times10^{-4}$, i.e.\
$\langle \tilde{X}\rangle\simeq3\times10^{14}\GeV$ for $c_{\cal O}$ of order unity, which corresponds
to small messenger Yukawa $\lambda_{\tilde{X}}\simeq3\times10^{-5}$ and small $k=(d+1)\langle {\tilde{X}}\rangle/M_{\rm Pl}\simeq10^{-3}$.
Because $T_d\simeq10\GeV$ lies well below the freeze-out temperature
$T_f\simeq m_\chi/20\simeq25\GeV$ and far above the BBN bound, neither the thermal history
nor photodissociation is affected. The two branches are then served by different source
masses: at $m_X=8\times10^{9}\GeV$ the multiplicity $1.3\times10^{-2}$ already exceeds
$B_{\max}=1.2\times10^{-2}$ for $m_\chi=100\TeV$, so the heavy branch needs
$m_X\gtrsim10^{10}\GeV$ whereas the light one needs $m_X\lesssim10^{10}\GeV$.

\subsection{Regeneration-safe dilution}
The difficulty of regeneration-safe dilution is a statement about the LSP multiplicity of the messenger decay,
not about entropy injection. The LSP multiplicity is determined by the structure of the messenger portal
operator of eq.~\eqref{eq:messenger-portal}.  Following from the $R$-parity
assignment, such operators predict essentially vanishing multiplicity. 

A further and more restrictive condition comes from the extra gauge
symmetry. As noted in eq.~(\ref{eq:gut-uniform-portal}), only specified GUT uniform messenger charge assignments are allowed by 
$U(1)_X$ gauge invariance of the messenger portal operator. 
Besides, in generic cases, either one Kahler operator or one superpotential operator is allowed by the $U(1)_X$ gauge symmetry. 
Scenario with GUT non-uniform anomaly-free $U(1)_X$ charge assignments had also been discussed in eq.~(\ref{eq:q-C}) and eq.~(\ref{eq:q-D}).

Given the allowed Kahler messenger portal for even R-parity scalar messenger, the decay vertex predicts $B_{\rm LSP}=0$, 
while the cascade it initiates adds the multiplicity $B_{\rm LSP}\simeq10^{-2}$ of section~\ref{sec:heavy-dilution}. 
The final yield is therefore $Y_\chi=(Y_{\rm th}+B_{\rm LSP}Y_X^{\rm before})/\Delta_d$, 
and the benchmark of eq.~\eqref{eq:heavy-benchmark}, $\Delta_d=2.8\times10^4$, returns $\Omega_\chi h^2=0.120$. 
Even if typical messenger portal is viable, two further conditions of different characters remain. First, the
entropy source $u=\frac34\Delta_d T_d\simeq82.8\,\GeV$ fixes the messenger
abundance through eq.~\eqref{eq:required-YX}, which gives $Y_X^{\rm before}\simeq8.3\times10^{-9}$ at 
$m_X=10^{10}\GeV$. A thermally produced messenger of that mass has $Y_X^{\rm th}\simeq(0.3-3)\times10^{-8}$, 
so the required $8.3\times10^{-9}$ lies inside this range and the source need not be generated non-thermally.
This presupposes a reheating temperature of order $m_X$ or above, so that
the messengers thermalise and freeze out with $Y_X^{\rm th}\propto m_X$, the
band reflecting the uncertainty in their annihilation rate. 

Because
$Y_X^{\rm before}\propto m_X^{-1}$ while $Y_X^{\rm th}\propto m_X$, the
two trends meet at $m_X\simeq(5-17)\times10^{9}\GeV$. Below that mass, the messenger source has to be 
generated non-thermally to reach 
the desired abundance $Y_X^{\rm before}$. 
 Second, the portal must be suppressed enough
to place the decay near the BBN bound. For example, at $m_X\sim10^8\GeV$ the same $T_d$ needs $|c_{\cal O}^{\rm eff}|\sim3\times10^{-5}$.

\subsection{Gravitinos, baryogenesis and the final dark-matter fraction}
\label{sec:gravitino}
\label{sec:cosmo-checks}

Gravitino plays different roles in minimal gauge mediation and deflected-AMSB.
In minimal gauge mediation, the gravitino is the lightest $R$-odd state and
is stable. In the deflected-AMSB realization, the gravitino is instead
heavier than the visible superpartners by a loop factor. So it is unstable and acts as an
{intermediate} state of the late-decay chain rather than the dark
matter candidate. 

In this work, the dark matter is the Higgsino. Gravitinos are produced
thermally rather than in messenger decay, because the goldstino channel
of eq.~\eqref{eq:messenger-goldstino-width} is suppressed by four powers
of $k$. Each gravitino decays gravitationally on its own timescale
[eq.~\eqref{eq:g3-2-width}], which lies after Higgsino freeze-out but
before or during the epoch relevant to BBN, and contributes one LSP
together with some entropy. A raw gravitino yield is not a final LSP yield when a subsequent source
dilutes it. The quantitative bounds depend on the gaugino spectrum,
reheating history and decay channels.

 The lightest messenger $\varphi_M$ is $R$-even and cannot decay into a gravitino, whose visible 
 decay proceeds through the Kahler portal with $B_{\rm LSP}=0$ and its
abundance is fixed by $Y_\varphi=u/m_X$ of eq.~\eqref{eq:required-YX}. The fermionic messenger $\psi_M$ is heavier. 
The dominant $\psi_M$ decay channel is into $\varphi_M$ plus a
gaugino, while its goldstino channel
$\psi_M\to\varphi_M+G_{3/2}$ carries the four $k$ powers of
eq.~\eqref{eq:messenger-goldstino-width}. Therefore,
\begin{equation}
 \frac{\Gamma_\lambda}{\Gamma(\chi_\phi\to\phi\lambda_a)}
 \simeq\frac{4\lambda^4k^4}{g^2}
 \left(\frac{F}{M_{\rm mess}^2\,\varepsilon}\right)^{2}\sim10^{-22},
 \label{eq:goldstino-suppressed}
\end{equation}
with $F=m_{3/2}M_{\rm Pl}$ and $\varepsilon=(1+d)F_\phi/M_{\rm mess}$ the
phase-space factor of eq.~\eqref{eq:messenger-gaugino-width}. So the
branching ratio into gravitinos is of order $10^{-17}$ for
$k\simeq8.2\times10^{-9}$ at $M_{\rm mess}=10^{10}\GeV$.

The fermionic messenger decays long before the scalar. For $N=1$,
$M_{\rm mess}=10^{10}\GeV$, $d=3$, $\lambda=0.4$ and
$F_\phi=3.5\times10^5\GeV$, eq.~\eqref{eq:messenger-gaugino-width}
gives $\Gamma_{\tilde G}\simeq0.16\GeV$, to be compared with
$H\simeq1.0\times10^{3}\GeV$ at $T\sim M_{\rm mess}$ for
$g_*=228.75$. Since $\Gamma_{\tilde G}/H\simeq1.6\times10^{-4}$
grows as $T^{-2}$ while the width is temperature independent,
$\Gamma_{\tilde G}=H$ is reached at $T\simeq1.3\times10^{8}\GeV$,
four orders of magnitude above the benchmark $T_d=4\MeV$. The
fermionic messenger therefore decays inside the hot plasma and carries no
late-time yield. Its decay
products, a gaugino and a messenger scalar, are injected into the plasma
at that temperature and re-thermalize. The scalar is still annihilating
faster than the expansion, so no excess scalar abundance survives.

 The gravitino abundance is
therefore the thermal one, 
\begin{equation}
 Y_{3/2}=10^{-12}\frac{T_R}{10^{10}\GeV}
 \left(\frac{m_{\widetilde g}}{1\TeV}\right)^2,
 \label{eq:tr-criterion}
\end{equation}
for reheating temperature $T_R$. 
Because the thermal Higgsino yield is diluted to
$Y_{\rm th}/\Delta_d=Y_{\rm req}$ by construction, the gravitino
contribution becomes comparable to it when
$Y_{3/2}^{\rm th}\simeq Y_{\rm th}=2.9\times10^{-11}$, that is when
\begin{equation}
 T_R\left(\frac{m_{\tilde g}}{1\TeV}\right)^{2}\simeq2.9\times10^{11}\GeV,
 \label{eq:tr-comparable}
\end{equation}
i.e.\ $T_R\simeq3.2\times10^{10}\GeV$ for $m_{\tilde g}=3\TeV$. At
$T_R=10^{10}\GeV$ the gravitinos carry the fraction
$Y_{3/2}^{\rm th}/Y_{\rm th}\simeq31\%$ of the thermal yield, i.e.\
$Y_{3/2}^{\rm th}/\Delta_d\simeq3.3\times10^{-16}$, about eight per cent
of $Y_{\rm req}$. This is an eight-per-cent correction to the yield of
eq.~\eqref{eq:heavy-benchmark}, within the accuracy of the benchmark.

The thermally produced gravitinos also release entropy just before the scalar messenger decays. 
Including
it multiplicatively, $\Delta_{\rm tot}=\Delta_{3/2}\Delta_d$ and
\begin{equation}
 Y_\chi=\frac{Y_{\rm th}}{\Delta_{3/2}\Delta_d}
 +\frac{Y_{3/2}^{\rm th}}{\Delta_d}.
 \label{eq:yield-with-gravitino-entropy}
\end{equation}
Here $\Delta_{3/2}=(1+\varepsilon_{3/2})^{3/4}$ is the dilution factor of
eq.~\eqref{eq:delta-def}, with
$\varepsilon_{3/2}\equiv4m_{3/2}Y_{3/2}^{\rm th}/(3T_{3/2})$ evaluated at the
pre-decay temperature. The gravitino entropy dilutes the
thermal yield but not the LSPs its own decay produces. Because the
gravitino number density is three orders of magnitude below that of the
scalar messenger, $\Delta_{3/2}-1\lesssim3\times10^{-4}$ even at the
$T_R$ where $Y_{3/2}^{\rm th}=Y_{\rm th}$, which is seven orders of
magnitude below $\Delta_d-1$. Hence $\Delta_{\rm tot}\simeq\Delta_d$, and
the gravitinos enter only as a non-thermal LSP source, through the second
term of the equation above.

With conserved $R$-parity, each gravitino decay yields exactly one LSP,
$B_{\rm LSP}^{(3/2)}=1$. The lifetime $\tau_{3/2}=9.7\times10^{-10}$--$7.7\times10^{-3}$~s (table~\ref{tab:gravitino-ref}) is shorter than the messenger lifetime
$0.093$ s, so gravitinos decay inside the messenger injection event
rather than after it, and their Higgsino products are diluted by the
subsequent entropy release. The thermal gravitino decays at
$T_{3/2}=31.5$--$104$ MeV for $m_{3/2}\sim 10^5 {\rm GeV}$, above the benchmark
$T_d=4$ MeV, and is diluted in the same way.

The only late-decaying messenger state is the scalar, which has
no gauge two-body channel and decays through the Kahler portal of
eq.~\eqref{eq:messenger-portal-decay} with $B_{\rm LSP}=0$. The gravitino
abundance is the thermal one,
$Y_{3/2}^{\rm th}\simeq9\times10^{-12}$ for
$T_R\simeq10^{10}\GeV$ and $m_{\tilde g}\simeq3\TeV$
(eq.~\eqref{eq:tr-criterion}), almost one third of $Y_{\rm th}$ and diluted in
the same way. It adds about eight per cent to the LSP yield, and we keep
it explicitly in the final abundance:
\begin{equation}
 Y_\chi=\frac{Y_{\rm th}+B_{\rm LSP}Y_X^{\rm before}}{\Delta_d}
+\frac{Y_{3/2}^{\rm th}}{\Delta_d}.
 \label{eq:yield-with-gravitino}
\end{equation}
With $\Delta_d\simeq2.8\times10^{4}$, $m_X=10^{10}\GeV$,
$Y_{\rm th}=2.9\times10^{-11}$ and $B_{\rm LSP}Y_X^{\rm before}=8.3\times10^{-11}$ and $Y_{3/2}^{\rm th}\simeq9\times10^{-12}$, this gives
$Y_\chi\simeq4.4\times10^{-15}$ and
$\Omega_\chi h^{2}\simeq0.12$: the entropy release, the injected
Higgsinos and the gravitino decays together reproduce the observed
abundance. 

Entropy production also dilutes a pre-existing baryon asymmetry. With no
later baryogenesis, $\eta_B^{\rm before}=\Delta_d\eta_B^{\rm today}$;
the benchmark $\Delta_d\simeq2.8\times10^{4}$ and
$\eta_B^{\rm today}=6.10\times10^{-10}$
require $\eta_B^{\rm before}\simeq1.7\times10^{-5}$. Alternatively, a
mechanism operating after the last entropy release would be needed.
Neither history is constructed here. Dark radiation, electromagnetic and
hadronic energy injection, and the resulting BBN constraints must also
be checked for the specified messenger decay model~\cite{Planck2018}.

For experimental comparisons, we use the \emph{final} abundance fraction
\begin{equation}
 f_\chi=\frac{\Omega_\chi h^2}{\Omega_{\rm DM}h^2}.
 \label{eq:final-fraction}
\end{equation}
The value $f_\chi\simeq0.217$ applies to the unmodified $500\GeV$ thermal
benchmark. If late production restores its abundance to $0.12$, then
$f_\chi=1$, not $0.217$. The same distinction must be maintained in all
direct-detection and solar-capture rates.

 
\section{Direct detection and solar-capture constraints}
\label{sec:dd}

\subsection{Inelastic scattering and the neutralino spectrum}
\label{sec:inelastic}

Inelastic Higgsino scattering probes the physical splitting defined in
eq.~\eqref{eq:splitting-definition} and the off-diagonal neutral-current
coupling obtained by diagonalizing eq.~\eqref{eq:matrix}. A small splitting
is not sufficient by itself: the light states must remain Higgsino-like,
and the extra singlets must not introduce a lighter LSP. The signed
bino and wino contributions in eq.~\eqref{eq:gaugino-split} may cancel,
but the additional term in eq.~\eqref{eq:singlet-splitting} must be retained
unless its decoupling has been demonstrated.

The event-motivated splitting interval quoted for a thermal-mass
Higgsino is approximately $340$--$360\keV$ in the $1.1 {\rm TeV}$ Higgsino
interpretations~\cite{Freese,FanReece,DuWang}. In the present model this
interval is reached only at the cancellation point identified in section~\ref{sec:model} above,
 where the splitting is a tuned rather than a generic output of
the singlet-extended $U(1)_X$ spectrum.

\subsection{Elastic scattering and abundance rescaling}
\label{sec:dd-limits}

For the per-nucleon spin-independent convention, write
\begin{equation}
 \sigma_{\rm SI}^{N}=
 \frac{4\mu_{\chi N}^2}{\pi}
 \left|\mathcal A_{\rm tree}^{N}+\mathcal A_{\rm loop}^{N}\right|^2,
 \label{eq:sigmaSI}
\end{equation}
where $\mu_{\chi N}$ is the reduced mass and the amplitudes include the
neutralino mixings and nucleon matrix elements. The tree-level scalar
exchange and electroweak loop terms can interfere. Their relative sign
must be computed in a common convention.

Spin-dependent scattering is treated in the same
convention, with the axial currents of the neutralino and of the nucleon,
\begin{equation}
 \sigma_{\rm SD}^N=\frac{3}{\pi}\mu_{\chi N}^2
 \left(\frac{a_Z\,g_A^{ZN}}{M_Z^2}\right)^2,
 \qquad
 \mathcal L\supset a_ZZ_\mu\bar\chi\gamma^\mu\gamma_5\chi
 +g_A^{ZN}Z_\mu\bar N\gamma^\mu\gamma_5N,
 \label{eq:sigmaSD}
\end{equation}
where $a_Z=(g/4c_W)X$ with $X$ the Higgsino asymmetry defined in
section~\ref{sec:elastic-alt} and $g_A^{ZN}\simeq0.5$. The values quoted
below follow eq.~\eqref{eq:sigmaSD} and are compared with the published
per-nucleon limits of appendix~\ref{app:phenodata}.

Limits quoted for a single dark-matter component can constrain
$f_\chi\sigma_{\rm SI}^{N}$. The final $f_\chi$ is defined in
eq.~\eqref{eq:final-fraction}. The reference LZ and XENONnT inputs are
collected in appendix~\ref{app:phenodata}~\cite{LZ2023,LZ42,Xenonnt2023,Xenonnt31}.
Their published best sensitivities occur at different WIMP masses and
must not be used as pointwise limits at $500\GeV$, $1.1\TeV$ or $100\TeV$
without the corresponding exclusion curves.

\subsection{Solar capture and the high-recoil interpretation}
\label{sec:solar}

Solar capture provides an independent test of inelastic Higgsino
scattering. For $m_\chi$ near $1.1\TeV$, the relevant solar analyses
impose exclusions extending to splittings of approximately
$506$--$566\keV$, depending on the treatment of scattering and
thermalization~\cite{Pospelov,Bose}.
 By these analyses we mean the standard equilibrium treatment, in
which the captured population reaches stationary balance, $C_\odot=\Gamma_A$,
so that the annihilation rate saturates at $\tfrac12C_\odot$ and the solar
neutrino flux takes its maximal value. Under those assumptions, the
$340$--$360\keV$ event-motivated interval has no overlap with the
surviving region. High-energy sideband data provide a further,
likelihood-dependent test~\cite{Rodd}. We therefore do not claim that the
minimal $1.1\TeV$ thermal-mass Higgsino interpretation explains the LZ candidate.

Changing the elastic amplitude can affect the solar thermalization
history, but an evasion requires a consistent capture, transport and
annihilation calculation. A suppressed elastic amplitude cannot simply
be substituted into an equilibrium capture bound. The exploratory
transport estimates retained in appendix~\ref{app:phenodata} are not a
validated solar likelihood and do not establish a surviving region.

At $m_\chi=100\TeV$, the smaller number density and altered capture and
annihilation dynamics require a new mass-dependent calculation. The
thermal-mass splitting limits cannot be transferred unchanged to this
regime. The relic-density construction of section~\ref{sec:relic} and
the possibility of a heavy-Higgsino event interpretation are therefore
separate questions~\cite{Langhoff}. A combined conclusion would require
mass-dependent recoil efficiencies, halo and nuclear inputs,
solar transport and the full neutralino spectrum~\cite{Baxter}.

\subsection{An elastic alternative with a bino- or singlino-like LSP}
\label{sec:elastic-alt}
The LZ candidate admits an alternative interpretation in terms of elastic dark matter scattering. 
The LSP can be
bino- or singlino-like and scatters through the diagonal axial $Z$ coupling,
which is set by the Higgsino asymmetry
$X_A=|N_{13}|^2-|N_{14}|^2$ rather than by the total Higgsino fraction. The
spin-dependent blind spot then does not coincide with the
spin-independent one, so the Higgs-mediated amplitude can be suppressed
while $X_A\sim10^{-2}$ survives~\cite{Roy:2026}. The scan of
fig.~\ref{fig:lsp-scan} contains this configuration: the bino and singlino
regions of that figure are the small-$\lambda$ and the $\lambda\simeq0.3$
corners of the parameter space. The benchmarks used in this
subsection therefore have a bino- or singlino-like LSP and belong to the
elastic branch, which are different parameter points from the inelastic
Higgsino benchmark of section~\ref{sec:inelastic}. Note that the two sets of
numbers should not be combined. The same deflection can host both branches:
at $d=5$ the LSP is Higgsino-like for $\mu=1.1\TeV$ and $F_\phi\simeq10^{6}\GeV$,
but bino-like for $\mu=700\GeV$ and $F_\phi\simeq10^{5}\GeV$, so $\mu$ and the
overall scale of $F_\phi$ select the branch rather than the deflection parameter $d$ alone.

\begin{figure}[tbp]
 \centering
 \includegraphics[width=0.62\textwidth]{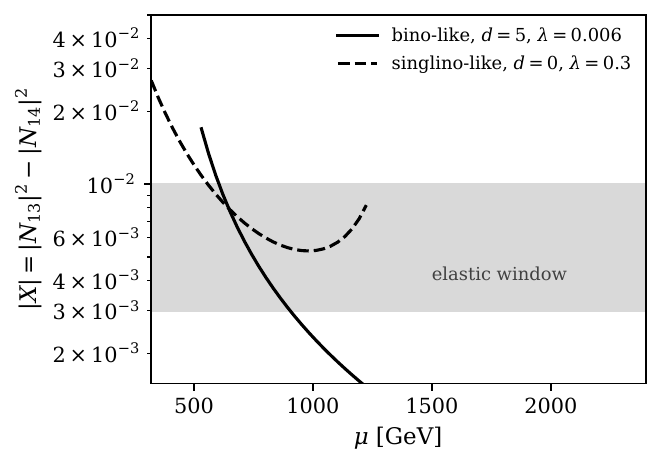}
 \caption{Higgsino asymmetry $|X_A|=|N_{13}|^2-|N_{14}|^2$ of the lightest
 neutralino along the two corridors of fig.~\ref{fig:lsp-scan} that give a
 bino-like (solid) or singlino-like (dashed) LSP, as a function of the
 Higgsino mass parameter. The shaded band marks the range relevant for the
 elastic interpretation of the high-recoil candidate~\cite{Roy:2026}. The
 curves are drawn only where the LSP is gaugino- or singlino-dominated.}
 \label{fig:elastic}
\end{figure}

Two points illustrate the two routes. For a bino-like LSP, $\lambda=0.006$,
$d=5$ and $F_\phi=1.27\times10^{5}\GeV$ give
$m_{\chi_1^0}=427\GeV$, a bino fraction of $0.99$ and
$X_A=-6.1\times10^{-3}$ at $\mu=700\GeV$. For a singlino-like LSP,
$\lambda=0.3$, $d=0$, $\mu=800\GeV$ and
$F_\phi=5.4\times10^{5}\GeV$ give
$m_{\chi_1^0}=480\GeV$, a singlino fraction of $0.92$ and
$X_A=5.5\times10^{-3}$. Both values agree with the range used in
ref.~\cite{Roy:2026}, and fig.~\ref{fig:elastic} shows that $X_A$ varies
smoothly with $\mu$ along each corridor: it falls from about
$3.7\times10^{-2}$ at $\mu\simeq450\GeV$ to $10^{-3}$ near
$\mu\simeq1.4\TeV$, crossing the elastic window $|X_A|\simeq10^{-2}$ without
tuning.

Points in the centre of that window are reached without
further tuning. Along the bino corridor, $\mu=600\GeV$ with $d=5$ and
$F_\phi=1.27\times10^{5}\GeV$ gives $m_{\chi_1^0}=426\GeV$,
$X_A=-1.03\times10^{-2}$ and
$\sigma_{\rm SD}=4.3\times10^{-42}\,\mathrm{cm}^2$; along the singlino
corridor, $\mu=550\GeV$ with $d=0$ and $F_\phi=5.4\times10^{5}\GeV$
gives $m_{\chi_1^0}=238\GeV$, $X_A=9.98\times10^{-3}$ and
$\sigma_{\rm SD}=4.0\times10^{-42}\,\mathrm{cm}^2$, in the same
normalization as the two benchmarks above. Both rates lie within a
factor of two of the present SD limit at the same mass, that is at the
sensitivity frontier rather than above it.

The origin of $X_A$ is model specific. Besides the usual gaugino--Higgsino
mixing, the $U(1)_X$ $D$ term couples the singlet and the two doublets
through $\xi_uv_u$ and $\xi_dv_d$, with $\xi_u=\xi_d=-0.2g_X$ but
$v_u\neq v_d$, and the singlino couples through
$\lambda v_u/\sqrt2$ and $\lambda v_d/\sqrt2$. The two Higgsino admixtures
are therefore unequal even though the two Higgs charges are equal, so $X_A$ is
neither protected nor tuned. Because $g_{YX}=M_{1X}=0$ for the uniform
content, no kinetic-mixing correction enters the blind-spot condition, which
is the MSSM one shifted by these two terms only.

At tree level, with $f_N=0.30$, the bino point gives
$\sigma_{\rm SI}\simeq7\times10^{-58}\,\mathrm{cm}^2$ and
$\sigma_{\rm SD}\simeq1.5\times10^{-42}\,\mathrm{cm}^2$, while the singlino
point gives $\sigma_{\rm SI}\simeq1\times10^{-49}\,\mathrm{cm}^2$
with $\sigma_{\rm SD}\simeq1.2\times10^{-42}\,\mathrm{cm}^2$.
The singlino benchmark lies close to its blind spot, so its
$\sigma_{\rm SI}$ falls below the corridor value at the same mass. 
Both benchmark rates lie below the present LZ sensitivity:
by ten to eleven orders of magnitude in the SI channel for the bino
point, and by about three for the singlino point, whose amplitude sits
nearer to its blind spot, but only by a factor of order two in the SD
channel at the same mass. These points are therefore consistent with
the candidate rather than a prediction of it. Whether the SD channel can
account for a single event is decided by the shape of the recoil
spectrum rather than by the rate: for the $|X_A|\simeq10^{-2}$ points
above, the exposure that yields one event above $240\keV$ yields of
order ten (bino) or seventy (singlino) events between $3$ and $70\keV$,
where LZ observes no excess. This ratio is independent of the cross
section and of the form-factor normalization, so a rate able to produce
the $248\keV$ event would need an SD cross section one to two orders of
magnitude above the values the model reaches at $|X_A|\simeq10^{-2}$, and
hence above the present SD limit. The SD channel can accommodate the
candidate but cannot explain it without such an enhancement.
The ratio $\sigma_{\rm SI}/\sigma_{\rm SD}$ is of order
$10^{-15}$ for the bino point and $10^{-8}$ for the singlino point, which is the
quantitative statement of the blind-spot separation.

\begin{figure}[tbp]
 \centering
 \includegraphics[width=0.92\textwidth]{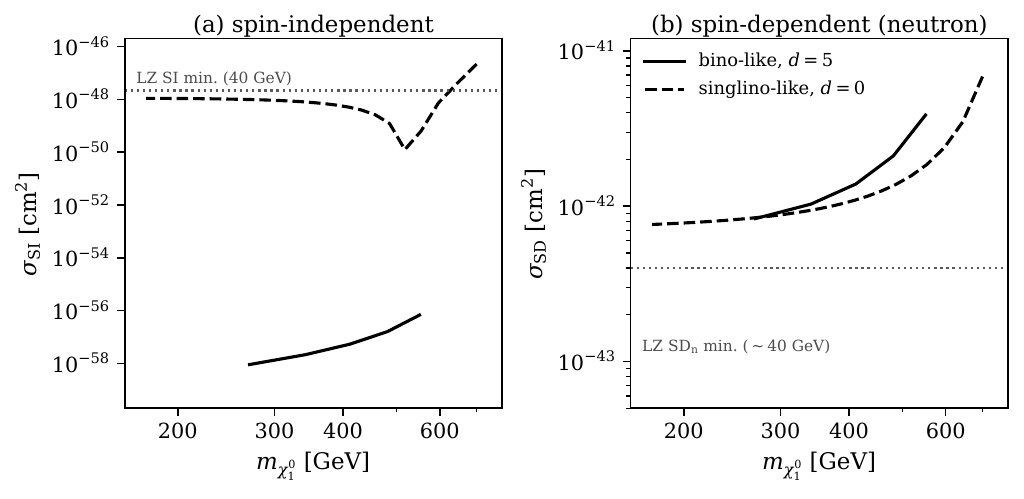}
 \caption{Spin-independent (a) and spin-dependent neutron (b) cross sections
 of the lightest neutralino along the two elastic corridors of
 fig.~\ref{fig:elastic}: the bino-like  curve with $F_\phi$ varied at
$d=5$, $\mu=700\GeV$ and $\lambda=0.006$, and the singlino-like
curve with $F_\phi$ varied at $d=0$, $\mu=800\GeV$ and
$\lambda=0.3$.
Horizontal dotted lines mark the minima of the LZ 90\% CL limits
($2.2\times10^{-48}\,\mathrm{cm}^2$ for SI and about
$4\times10^{-43}\,\mathrm{cm}^2$ for SD, both at their best mass near
$40\GeV$). The dip of the
singlino SI curve is the blind-spot crossing, where the Higgs-mediated
amplitude changes sign.}
 \label{fig:sigma-m}
\end{figure}

The two channels behave differently along the corridors
(fig.~\ref{fig:sigma-m}). Over $m_{\chi_1^0}\simeq160$--$600\GeV$ the
bino-like SI cross section stays between $10^{-58}$ and
$2\times10^{-56}\,\mathrm{cm}^2$, while its SD cross section remains in the
$10^{-43}$--$10^{-42}\,\mathrm{cm}^2$ band. The singlino-like SI cross
section is larger by eight to ten orders of magnitude relative to the bino
corridor, of order $10^{-48}\,\mathrm{cm}^2$, and dips by about two orders
at the blind spot near $530\GeV$, where $g_{h\chi\chi}$ changes sign, while its
SD cross section tracks the bino one. In both corridors, SD is the more
accessible channel, as required by the axial-$Z$ interpretation, while SI
stays below the LZ limit at the same mass.

Two differences from the NMSSM realization are worth noting. First, there
is no $\kappa S^3$ term: the CP-odd component of $S$ is eaten by the $Z'$,
so the model has no light singlet-like pseudoscalar $a_S$. Unlike in the NMSSM,
the relevant relic abundance of these parameter points therefore cannot be obtained resonantly,
through $\chi\chi\to a_S$ with $m_{a_S}$ just above the $2m_\chi$ threshold.
Second, the $U(1)_X$ $D$-term potential,
$\frac{g_X^2}{2}(q_u|H_u|^2+q_d|H_d|^2+\dots)^2$, contains the quartic
cross term $g_X^2q_uq_d|H_u|^2|H_d|^2$. This is the analogue of the NMSSM
tree-level term $\lambda^2|H_u|^2|H_d|^2$ that relieves the low-$\tan\beta$
Higgs-mass tension in ref.~\cite{Roy:2026}.

The limitations are the same as in the rest of this section: $\mu$ is an
input, the cross sections are tree level, the relic abundance of these
gaugino- and singlino-like points has not been computed, and the
blind-spot condition must be verified numerically point by point rather
than assumed. The two branches nevertheless predict different recoil
spectra away from the candidate, so the same data set can separate them.

\section{Conclusions}
\label{sec:discussion}
\label{sec:novelty}
\label{sec:cost-ledger}
Motivated by the recent LZ high-recoil event, we have studied the Higgsino dark matter scenario in a construction that combines the charged-singlet solution of the $\mu$ problem with deflected anomaly mediation in a $U(1)_X$ framework. The gauge symmetry carries both roles: it supplies the
additional gauge sector, forbids the Higgs bilinear and the cubic $S^3$,
and regenerate the $\mu$-term after the singlet acquires a VEV that breaks the $U(1)_X$.

Because $S$ is charged under $U(1)_X$, the $\widetilde B_X$ gaugino does not decouple
and the active $6\times 6$ neutralino mass matrix also includes the $\widetilde B_X$ and singlino components, in addition to the ordinary bino-wino-Higgsino ones. Its neutral Higgsino splitting contains the signed
gaugino contribution, a singlino term fixed by
eq.~\eqref{eq:singlino-ratio}, and an Abelian gaugino correction. Full diagonalization,
threshold matching and radiative corrections remain necessary for precision benchmarks.

The predictive deflected AMSB scenario can predict a light dark matter through
proper deflection parameters and messenger content. The opposite-sign
electroweak-gaugino branch allows a suppressed gaugino contribution to
the splitting, while the supersymmetry-breaking scale and singlet
inputs determine whether a light or heavy Higgsino hierarchy can be
realized.

The presence of a messenger sector in deflected AMSB can have important cosmological consequences, which shows key differences from that of ordinary GMSB. The cosmological consequences of the messenger sector in deflected AMSB have received little attention so far, yet they affect the thermal history of the dark matter substantially.
In deflected AMSB, the gravitino is heavier than the visible superpartners
and unstable, so the dark matter is the Higgsino. The messenger sector
provides both non-thermal Higgsino production and entropy injection.
Messenger decays can replenish the underabundant $500\GeV$ thermal
Higgsino through direct and cascade channels. A negligible or balanced
source can preserve the $1.1\TeV$ thermal candidate and a dominating
messenger can dilute the overabundant $100\TeV$ thermal population by entropy injection. The lightest messenger is normally an $R$-even scalar, which can decay via the
Planck-suppressed operators of eq.~\eqref{eq:messenger-portal} and only the
Kahler class gives it two-body decays into Standard-Model fermions with $B_{\rm LSP}=0$ at leading order.
For points whose pre-dilution yield exceeds $Y_{\rm req}$, entropy dilution
alone reproduces $\Omega_\chi h^2=0.12$; points that undershoot it cannot be
recovered, since the $R$-even messenger carries $B_{\rm LSP}=0$. The
superpotential class leads to the same conclusion: its three-body channels
into three Standard-Model fermions are $R$-parity allowed and carry
$B_{\rm LSP}=0$ at leading order, channels with a single superpartner are $R$-parity
forbidden, and two-LSP final states are not generated by the operators of
eq.~\eqref{eq:messenger-portal}. These statements however concern the decay
vertex. As argued in section~\ref{sec:heavy-dilution}, the debris that the
portal injects can still pair-produce superpartners, so the total $B_{\rm LSP}\lesssim 10^{-2}$. 

The same spectrum also accommodates the elastic interpretation of the high-recoil candidate. 
For a bino- or singlino-like LSP, the diagonal axial $Z$ coupling is the Higgsino asymmetry $X_A=|N_{13}|^2-|N_{14}|^2$. 
The $U(1)_X$ $D$-term together with the singlino admixtures moves it away from the MSSM blind spot, 
so the spin-dependent blind spot does not coincide with the spin-independent one. 
The value $|X_A|\sim10^{-2}$ required by a $248\keV$ recoil can be obtained without tuning. 
The bino-like point then has $\sigma_{\rm SI}\simeq7\times10^{-58}\,\mathrm{cm}^2$, ten to eleven orders below the LZ limit, 
and $\sigma_{\rm SD}\simeq1.5\times10^{-42}\,\mathrm{cm}^2$, within a factor of two of the present limit at the same mass. 
The singlino-like point has a similar $\sigma_{\rm SD}$ with a larger $\sigma_{\rm SI}$. 
The resulting spin-dependent rate then lies close to the present experimental sensitivity.

Direct detection and solar capture constrain the final abundance and
the physical neutralino couplings, not the unmodified thermal abundance.
The high-recoil candidate remains a phenomenological motivation rather
than evidence for this construction, and the stated thermal-mass solar
constraints cannot be evaded by an entropy argument alone. A completed
analysis must combine the $U(1)_X$ singlet spectrum, a specified messenger
decay calculation and the experimental likelihoods. 

\acknowledgments
 This work was supported by the
Natural Science Foundation of China under grant numbers 12335005 and 12075213; by the Key Research Project of Henan Education Department for colleges and universities under grant number 21A140025.
Deepseek-assisted tools were used to edit the language and organize the numerical calculations.


\appendix

%
%

\section{Soft supersymmetry-breaking parameters}
\label{app:soft}

This appendix gives the complete deflected-anomaly-mediated soft parameters
 of the model of sec.~\ref{sec:model}, with the uniform charge assignment
 \eqref{eq:gut-uniform-charges} and the light spectrum of
 eq.~\eqref{eq:wlow1}. Expressions for the
soft supersymmetry breaking scalar masses are
decomposed into the pure deflected-AMSB piece $\delta_d$, an
anomaly--gauge interference piece $\delta_I$ and a pure gauge-mediation
 piece $\delta_G$ --- follows appendix~A of ref.~\cite{damsbsoft}. The form of the soft SUSY breaking parameters predicted by deflected AMSB are very different from those of the ordinary NMSSM, including the extra $U(1)_X$ factor, the
Abelian kinetic mixing $g_{YX}$, the absence of $\kappa S^3$, the
$X$-charged singlet $S$ et.

Using the order parameter $F_\phi\equiv F/M_{\rm Pl}$ and the boundary
condition of eq.~\eqref{eq:gaugino}, the gaugino masses can be given by
\begin{equation}
 M_i(M_{\rm mess})=-\frac{F_\phi\,g_i^2(M_{\rm mess})}{16\pi^2}\,(b_i-Nd),
 \label{eq:soft-gaugino-bound}
\end{equation}
where $N$ is the number of messenger pairs and $d$ the dimensionless
deflection parameter. We write
\begin{equation}
 \alpha_a\equiv\frac{g_a^2}{4\pi},\qquad
\alpha_1=\tfrac53\alpha_Y ,
 \label{eq:soft-alpha}
\end{equation}
with $g_Y$ the \emph{physical} hypercharge coupling and $g_1$ the
GUT-normalized one.

Anomalous dimensions, Yukawa beta functions, $A$-terms and scalar masses are
defined by
\begin{equation}
 \gamma_i=\frac12\frac{{\rm d}\ln Z_i}{{\rm d}\ln Q},\qquad
 m^2_{i,{\rm AMSB}}=\frac{F_\phi^2}{2}\,\dot\gamma_i,\qquad
 A_y=-F_C\frac{\beta_y}{y}=+\frac{F_\phi}{16\pi^2}\,\widetilde G_y ,
 \qquad \dot\gamma\equiv\frac{{\rm d}\gamma}{{\rm d}\ln Q},
 \label{eq:soft-defs}
\end{equation}
and we abbreviate
\begin{equation}
 \widetilde G_y\;\equiv\;16\pi^2\,\frac{\beta_y}{y}
 \;=\;\sum_{i\,\in\, y}16\pi^2\gamma_i ,
 \qquad
 16\pi^2\gamma_i=\sum_y c_i(y)\,y^2-2C_3(R_i)g_3^2-2C_2(R_i)g_2^2-2\Xi_i .
 \label{eq:soft-gtilde-def}
\end{equation}
The coefficients $c_i(y)$ are fixed by the index structure of
eq.~\eqref{eq:wlow1}.

\paragraph{Abelian charge vectors.}

Write $D_\mu=\partial_\mu-i\,q_i^{T}{\cal G}\,A_\mu$ with
$A_\mu=(B_\mu,X_\mu)^T$ and
\begin{equation}
 {\cal G}=\begin{pmatrix} g_Y & g_{YX}\\ 0 & g_X\end{pmatrix},
 \qquad
 {\cal H}\equiv{\cal G}{\cal G}^T
 =\begin{pmatrix} g_Y^2+g_{YX}^2 & g_{YX}g_X\\
                  g_{YX}g_X & g_X^2\end{pmatrix}.
 \label{eq:soft-Gcal}
\end{equation}
This triangular form follows from removing the kinetic mixing.
Starting from 
\beqa \mathcal L\supset-\tfrac14B_{\mu\nu}B^{\mu\nu}
-\tfrac14X_{\mu\nu}X^{\mu\nu}-\tfrac{\varepsilon}{2}B_{\mu\nu}X^{\mu\nu}~,\eeqa
the shifts 
\beqas 
B_\mu\to B_\mu+\varepsilon X_\mu~,~~~X_\mu\to X_\mu/\sqrt{1-\varepsilon^2}
\eeqas  
bring the gauge kinetic terms canonical while turning the $X$ coupling into
$g_Xq_X-\varepsilon g_YY$. Comparison with the matrix above then
identifies $g_{YX}\simeq-\varepsilon g_Y$ at leading order in
$\varepsilon$. Only the combination $\mathcal H=\mathcal G\mathcal
G^{\,T}$ enters gauge-invariant observables, so the choice among
triangular bases is a convention. The coupling $g_{YX}$ runs according to the
matrix equation 
\beqa 16\pi^2\dot{\cal G}={\cal G}\,{\cal G}^TB{\cal G}~.\eeqa

Field $i$ couples to $B_\mu$ with $g_YY_i$ and to $X_\mu$ with
$g_{YX}Y_i+g_Xq_{X,i}$. The latter is exactly the $\xi_i$ of the neutralino
matrix \eqref{eq:matrix}, since $\xi_u=g_Xq_{H_u}+g_{YX}/2$ and
$\xi_d=g_Xq_{H_d}-g_{YX}/2$. Define the two Abelian invariants
\begin{align}
 \Xi_i &\equiv q_i^T{\cal H}q_i
   =(g_Y^2+g_{YX}^2)Y_i^2+2g_{YX}g_XY_iq_{X,i}+g_X^2q_{X,i}^2
   =g_Y^2Y_i^2+\bigl(g_{YX}Y_i+g_Xq_{X,i}\bigr)^2,
   \label{eq:soft-Xi-def}\\[1mm]
 u_i &\equiv {\cal H}q_i
   =\binom{(g_Y^2+g_{YX}^2)Y_i+g_{YX}g_Xq_{X,i}}
          {g_{YX}g_XY_i+g_X^2q_{X,i}}
   \equiv\binom{u_{i,Y}}{u_{i,X}} .
   \label{eq:soft-u-def}
\end{align}
$\Xi_i$ controls the one-loop anomalous dimensions, $u_i$ the two-loop
(deflected) scalar masses. We also use
\begin{equation}
 {\cal Y}\equiv g_Y^2+g_{YX}^2,\qquad
 {\cal Z}\equiv g_{YX}g_X,\qquad
 {\cal X}\equiv g_X^2 .
 \label{eq:soft-YZX}
\end{equation}

The messenger sector consists of $N$ \emph{complete} $SU(5)$
$\mathbf5+\overline{\mathbf5}$ pairs,
\begin{equation}
 \mathbf5=(\mathbf3,\mathbf1)_{Y=-1/3}\oplus(\mathbf1,\mathbf2)_{Y=+1/2},
 \qquad
 \overline{\mathbf5}=(\overline{\mathbf3},\mathbf1)_{Y=+1/3}
                     \oplus(\mathbf1,\mathbf2)_{Y=-1/2},
\end{equation}
 with the uniform $U(1)_X$ charge assignment of
 eq.~\eqref{eq:gut-uniform-charges}, for which both components of a multiplet
 carry the same charge,
\begin{equation}
 q_X(\mathbf5)=q_M,\qquad
 q_X(\overline{\mathbf5})=-q_M,
 \label{eq:soft-mess-qx}
\end{equation}
 normalised to $q_M=2q=2$ together with the visible charges. The value is
 fixed by the decay portal: the K\"ahler operators of
 eq.~\eqref{eq:messenger-portal} are invariant only for
 $q_M=-2q$ for $\bar5_M^\dagger10_F^{\,2}$ and $q_M=+2q$
 for $5_M^\dagger\bar5_F10_F$, while the two superpotential operators would need
 $q_M=3q$ and $q_M=-q$, respectively.
Completeness of the $SU(5)$ multiplets makes $\Delta b_i=N$
{universal} over the three SM factors.
The $X$-charge assignment \eqref{eq:soft-mess-qx} determines both
 $\Delta b_X=2N(3q_M^2+2q_M^2)=40N$ and
 $\Delta B_{YX}=N(3q_M-3q_M+2q_M-2q_M)=0$. For the uniform assignment, the
 messenger pair generates no mixed trace, and no kinetic mixing is produced
 at the messenger threshold. The above-threshold
coefficients are
\begin{equation}
 b^{\rm UV}=b+\Delta b=\Bigl(\frac{37}{5}+N,\;1+N,\;-1+N,\;128+\Delta b_X\Bigr).
 \label{eq:soft-bUV}
\end{equation}
\subsection{Gaugino masses}
\label{app:soft-gaugino}
 With $(b_1,b_2,b_3,b_X)=(37/5,1,-1,128)$,
\begin{equation}
 M_2(M_{\rm mess})=-\frac{F_\phi g_2^2}{16\pi^2}\,(1-Nd),\qquad
 M_3(M_{\rm mess})=-\frac{F_\phi g_3^2}{16\pi^2}\,(-1-Nd).
 \label{eq:soft-M23}
\end{equation}
Set the Abelian matrix
\begin{equation}
 B'\equiv B_{(Y,X)}-d\,\Delta B
 =\begin{pmatrix}\frac{37}{3}-\frac{5Nd}{3}&0\\[1mm]0&128-\Delta b_X d\end{pmatrix},
 \label{eq:soft-Bprime}
\end{equation}
and define $\Pi\equiv{\cal G}^TB'{\cal G}$, then
\begin{equation}
 M_{ab}=-\frac{F_\phi}{16\pi^2}\,\Pi_{ab}.
 \qquad a,b\in\{B,X\},
 \label{eq:soft-abelian-matrix}
\end{equation}

Hence
\beqa
  M_{YY}\equiv M_1&=&-\frac{F_\phi g_Y^2}{16\pi^2}
      \Bigl(\frac{37}{3}-\frac{5Nd}{3}\Bigr)=-(F_\phi g_1^2/16\pi^2)(b_1-Nd),\\[1mm]
  M_{1X}&=&-\frac{F_\phi g_Y}{16\pi^2}\,
      g_{YX}\Bigl(\frac{37}{3}-\frac{5Nd}{3}\Bigr),\\[1mm]
  M_{XX}&=&-\frac{F_\phi}{16\pi^2}
      \Bigl[g_{YX}^2\Bigl(\frac{37}{3}-\frac{5Nd}{3}\Bigr)
            +g_X^2\Bigl(128-\Delta b_X d\Bigr)\Bigr].
  \label{eq:soft-Mabelian}
 \eeqa
\subsection{Trilinear couplings}
\label{app:soft-gamma}
The anomalous dimensions for various superfields are
\begin{align}
 16\pi^2\gamma_{Q_a}   &= y_{u,a}^2+y_{d,a}^2-\tfrac83g_3^2-\tfrac32g_2^2-2\Xi_Q,
 &
 16\pi^2\gamma_{u^c_a} &= 2y_{u,a}^2-\tfrac83g_3^2-2\Xi_{u^c},
 \nonumber\\
 16\pi^2\gamma_{d^c_a} &= 2y_{d,a}^2-\tfrac83g_3^2-2\Xi_{d^c},
 &
 16\pi^2\gamma_{L_a}   &= y_{e,a}^2+y_{\nu,a}^2-\tfrac32g_2^2-2\Xi_L,
 \nonumber\\
 16\pi^2\gamma_{e^c_a} &= 2y_{e,a}^2-2\Xi_{e^c},
 &
 16\pi^2\gamma_{N^c_a} &= 2y_{\nu,a}^2-2\Xi_{N^c},
 \nonumber\\
 16\pi^2\gamma_{H_u}   &= 3y_t^2+y_\nu^2+\lambda^2-\tfrac32g_2^2-2\Xi_{H_u},
 &
 16\pi^2\gamma_{H_d}   &= 3y_b^2+y_\tau^2+\lambda^2-\tfrac32g_2^2-2\Xi_{H_d},
 \nonumber\\
 16\pi^2\gamma_{S}     &= 2\lambda^2+3\sum_{i=1}^2y_{D_i}^2-2\Xi_S,
 &
 16\pi^2\gamma_{D_i}   &= y_{D_i}^2-\tfrac83g_3^2-2\Xi_{D_i},
 \nonumber\\
 16\pi^2\gamma_{D^c_i} &= y_{D_i}^2-\tfrac83g_3^2-2\Xi_{D^c_i}.
 \label{eq:soft-gammas}
\end{align}
with 
{\small
\begin{equation}
 \begin{array}{rcl@{\qquad}rcl}
 \Xi_Q   &=&\dfrac{{\cal Y}+12{\cal Z}+36{\cal X}}{36},
   & \Xi_S   &=& 16{\cal X},\\[2mm]
 \Xi_{u^c}&=&\dfrac{4{\cal Y}-12{\cal Z}+9{\cal X}}{9},
   & \Xi_{D_i}   &=&\dfrac{{\cal Y}+12{\cal Z}+36{\cal X}}{9},\\[2mm]
 \Xi_{d^c}&=&\dfrac{{\cal Y}+6{\cal Z}+9{\cal X}}{9},
   & \Xi_{D^c_i}&=&\dfrac{{\cal Y}-12{\cal Z}+36{\cal X}}{9},\\[2mm]
 \Xi_L   &=&\dfrac{{\cal Y}-4{\cal Z}+4{\cal X}}{4},
   & \Xi_{N^c} &=& {\cal X},\\[2mm]
 \Xi_{e^c}&=&{\cal Y}+2{\cal Z}+{\cal X},
   & \Xi_{H_u}&=&\dfrac{{\cal Y}-8{\cal Z}+16{\cal X}}{4},\\[2mm]
 \Xi_{H_d}&=&\dfrac{{\cal Y}+8{\cal Z}+16{\cal X}}{4}.
 \end{array}
 \label{eq:soft-Xi-table}
\end{equation}}
The anomalous dimensions for the Yukawa couplings are thus
\begin{align}
 \widetilde G_{y_t}
  &= \lambda^2+6y_t^2+y_b^2+y_\nu^2-\frac{16}{3}g_3^2-3g_2^2
     -\frac{13}{9}{\cal Y}+6{\cal Z}-12{\cal X},
     \label{eq:soft-Gt}\\[1mm]
 \widetilde G_{y_b}
  &= \lambda^2+y_t^2+6y_b^2+y_\tau^2-\frac{16}{3}g_3^2-3g_2^2
     -\frac{7}{9}{\cal Y}-6{\cal Z}-12{\cal X},
     \label{eq:soft-Gb}\\[1mm]
 \widetilde G_{y_\tau}
  &= \lambda^2+y_\nu^2+3y_b^2+4y_\tau^2-3g_2^2
     -3{\cal Y}-6{\cal Z}-12{\cal X},
     \label{eq:soft-Gtau}\\[1mm]
 \widetilde G_{y_\nu}
  &= \lambda^2+y_\tau^2+3y_t^2+4y_\nu^2-3g_2^2
     -{\cal Y}+6{\cal Z}-12{\cal X},
     \label{eq:soft-Gnu}\\[1mm]
 \widetilde G_{\lambda}
  &= 4\lambda^2+3y_t^2+3y_b^2+y_\tau^2+y_\nu^2+3\sum_{i=1}^2y_{D_i}^2
     -3g_2^2-{\cal Y}+0\cdot{\cal Z}-48{\cal X},
     \label{eq:soft-Glam}\\[1mm]
 \widetilde G_{y_{D_i}}
  &= 2\lambda^2+3\sum_{j=1}^{2}y_{D_j}^2+2y_{D_i}^2-\frac{16}{3}g_3^2
     -\frac{4}{9}{\cal Y}+0\cdot{\cal Z}-48{\cal X}.
     \label{eq:soft-GyD}
\end{align}
We adopt diagonal Dirac Yukawa couplings for neutrino. That coupling is, however, numerically negligible. The right-handed neutrinos obtain their mass only from the Dirac term $y_\nu N^cLH_u$, so the observed scale $m_\nu\simeq y_\nu v_u/\sqrt2\lesssim0.1\ev$
implies $y_\nu\lesssim10^{-12}$. Every $y_\nu$ entry can therefore be dropped in the soft-breaking expressions: the $y_\nu^2$ terms of $\widetilde G_{y_t}$, $\widetilde G_{y_\tau}$, $\widetilde G_{y_\nu}$ and $\widetilde G_\lambda$ are set to zero, and $\widetilde G_{y_\nu}$ reduces to its gauge part.

Because of the $U(1)_X$ charge of $S$, $\widetilde G_\lambda$ depend on $g_X$ at one loop, which is absent in the NMSSM with a neutral singlet.

The soft SUSY breaking trilinear couplings $T_y=yA_y$ are given by
\begin{equation}
 A_y=\frac{F_\phi}{16\pi^2}\Bigl[\widetilde G_y-d\,\Delta\widetilde G_y\Bigr],
 \qquad ,
 \label{eq:soft-Ageneral}
\end{equation}
where $\Delta\widetilde G_y$ is the discontinuity of $16\pi^2\beta_y/y$
across the messenger threshold. Because no messenger-matter interactions are present,
\begin{equation}
 \Delta\widetilde G_y=0\quad\text{for all }y\in
 \{y_t,y_b,y_\tau,y_\nu,\lambda,y_{D_i}\}.
 \label{eq:soft-DGt-zero}
\end{equation}
So, we have 
\begin{equation}
 A_t=\frac{F_\phi}{16\pi^2}\widetilde G_{y_t},\quad
 A_b=\frac{F_\phi}{16\pi^2}\widetilde G_{y_b},\quad
 A_\tau=\frac{F_\phi}{16\pi^2}\widetilde G_{y_\tau},
 \label{eq:soft-Alight}
\end{equation}
\begin{equation}
 A_\nu=\frac{F_\phi}{16\pi^2}\widetilde G_{y_\nu}, \quad A_\lambda=\frac{F_\phi}{16\pi^2}\widetilde G_{\lambda}, \quad A_{y_{D_i}}=\frac{F_\phi}{16\pi^2}\widetilde G_{y_{D_i}} .
 \label{eq:soft-Aexotic}
\end{equation}
There is no $A_\kappa$, as the operator $\kappa S^3$ is forbidden by $U(1)_X$.

\subsubsection{Scalar soft masses: $\delta_d$}
\label{app:soft-scalar}

We parameterize the soft scalar masses in the form of ref.~\cite{damsbsoft}
\begin{equation}
 m^2_{\tilde f_i}=\delta_d^{\,i}+\delta_I^{\,i}+\delta_G^{\,i},
 \label{eq:soft-decomp}
\end{equation}
with $\delta_d$ the $d$-independent pure anomaly-mediation contributions,
$\delta_I$ the collection of everything linear in $d$, and $\delta_G$ everything
quadratic in $d$. They split uniquely by their power in the
deflection parameter  
\begin{equation}
 \delta_d\equiv\delta_{\rm AMSB}\;(d^0),\qquad
 \delta_I=\delta_I^{\rm gauge}+\delta_I^{\rm Yukawa}\;(d^1),\qquad
 \delta_G=\delta_G^{\rm gauge}+\delta_G^{\rm Yukawa}\;(d^2).
 \label{eq:soft-split-def}
\end{equation}

Deflection can enter the scalar masses through
$b_a\to b_a-(d^2+2d)\Delta b_a$, i.e.
\begin{equation}
 G_a\equiv(d^2+2d)\Delta b_a-b_a .
 \label{eq:soft-Ga-def}
\end{equation}
For the Abelian sector the object that runs is the matrix $B_{(Y,X)}$, so the
deflection is the matrix replacement
\begin{equation}
 B_d\equiv B_{(Y,X)}-(d^2+2d)\Delta B,\qquad
 G_Y\equiv-(B_d)_{YY},\quad G_{YX}\equiv-(B_d)_{YX},\quad
 G_X\equiv-(B_d)_{XX},
 \label{eq:soft-Bd}
\end{equation}
 which with eq.~\eqref{eq:soft-Bprime} gives 
\begin{equation}
 \begin{aligned}
 G_1&=N(d^2+2d)-\frac{37}{5},&\qquad
 G_2&=N(d^2+2d)-1,\\[1mm]
 G_3&=N(d^2+2d)+1,&\qquad
 G_{YX}&=0,\\[1mm]
 G_Y&=\frac{5N}{3}(d^2+2d)-\frac{37}{3}=\frac53G_1,&\qquad
 G_X&=\Delta b_X(d^2+2d)-128.
 \end{aligned}
 \label{eq:soft-Gall}
\end{equation}
 Note that $G_{YX}=0$ identically. For the uniform content, the mixed trace $B_{YX}$
 vanishes and the messenger pair adds $\Delta B_{YX}=0$. So, no
 kinetic-mixing contribution to any scalar mass is generated at this order.
There is no messenger--matter Yukawa coupling. Consequently, the one-loop anomalous dimensions for the matter superfields are {continuous} across the messenger threshold.

The first part of soft scalar mass is given by
\begin{equation}
 \begin{aligned}
 \delta_d^{\,i}&=\frac{F_\phi^2}{(16\pi^2)^2}
 \Bigl\{\,2C_3(R_i)\,G_3\,g_3^4+2C_2(R_i)\,G_2\,g_2^4\\
 &\quad+2\bigl[G_Yu_{i,Y}^2+2G_{YX}u_{i,Y}u_{i,X}+G_Xu_{i,X}^2\bigr]
 +\sum_y c_i(y)\,y^2\widetilde G_y\Bigr\}
 \end{aligned}
 \label{eq:soft-deltad-master}
\end{equation}
with $u_i$ from eq.~\eqref{eq:soft-u-def}, $c_i(y)$ from
 eq.~\eqref{eq:soft-gammas}, $G_a$ from eq.~\eqref{eq:soft-Gall} and
 $\widetilde G_y$ from eqs.~\eqref{eq:soft-Gt}--\eqref{eq:soft-GyD}. The first
two terms are the analogues of eq.~(A.6) of ref.~\cite{damsbsoft}. The third term
is new ($U(1)_X$ plus kinetic mixing) while the fourth term is the two-loop Yukawa term
of eqs.~(A.6), (A.8) and (A.9).

 In the diagonal-kinetic
limit $g_{YX}=0$, where $u_{i,Y}=g_Y^2Y_i$ and $u_{i,X}=g_X^2q_{X,i}$,
\begin{equation}
 \begin{aligned}
 \delta_d^{\,i}\Big|_{g_{YX}=0}&=\frac{F_\phi^2}{(16\pi^2)^2}
 \Bigl\{w_3^{(i)}G_3g_3^4+w_2^{(i)}G_2g_2^4
       +w_Y^{(i)}G_Yg_Y^4\\
 &\quad+w_{YX}^{(i)}G_{YX}g_Y^2g_X^2
       +w_X^{(i)}G_Xg_X^4+\sum_y c_i(y)y^2\widetilde G_y\Bigr\}
 \end{aligned}
 \label{eq:soft-deltad-diag}
\end{equation}
with completely explicit weights
{\scriptsize
\begin{equation}
 \begin{array}{c|rrrrr|l}
  & w_3=2C_3 & w_2=2C_2 & w_Y=2Y^2 & w_{YX}=4Yq_X & w_X=2q_X^2
  & \sum_y c_i(y)y^2\widetilde G_y\\ \hline
 \widetilde Q_a      & \frac83 & \frac32 & \frac1{18} & \frac23 & 2
   & y_{u,a}^2\widetilde G_{y_t}+y_{d,a}^2\widetilde G_{y_b}\\
 \widetilde u^c_a    & \frac83 & 0 & \frac89 & -\frac83 & 2
   & 2y_{u,a}^2\widetilde G_{y_t}\\
 \widetilde d^c_a    & \frac83 & 0 & \frac29 & \frac43 & 2
   & 2y_{d,a}^2\widetilde G_{y_b}\\
 \widetilde L_a      & 0 & \frac32 & \frac12 & -2 & 2
   & y_{e,a}^2\widetilde G_{y_\tau}+y_{\nu,a}^2\widetilde G_{y_\nu}\\
 \widetilde e^c_a    & 0 & 0 & 2 & 4 & 2
   & 2y_{e,a}^2\widetilde G_{y_\tau}\\
 \widetilde N^c_a    & 0 & 0 & 0 & 0 & 2
   & 2y_{\nu,a}^2\widetilde G_{y_\nu}\\
 \widetilde H_u      & 0 & \frac32 & \frac12 & -4 & 8
   & 3y_t^2\widetilde G_{y_t}+y_\nu^2\widetilde G_{y_\nu}
     +\lambda^2\widetilde G_\lambda\\
 \widetilde H_d      & 0 & \frac32 & \frac12 & +4 & 8
   & 3y_b^2\widetilde G_{y_b}+y_\tau^2\widetilde G_{y_\tau}
     +\lambda^2\widetilde G_\lambda\\
 \widetilde S        & 0 & 0 & 0 & 0 & 32
   & 2\lambda^2\widetilde G_\lambda
     +3\textstyle\sum_iy_{D_i}^2\widetilde G_{y_{D_i}}\\
 \widetilde D_i      & \frac83 & 0 & \frac29 & \frac83 & 8
   & y_{D_i}^2\widetilde G_{y_{D_i}}\\
 \widetilde D^c_i    & \frac83 & 0 & \frac29 & -\frac83 & 8
   & y_{D_i}^2\widetilde G_{y_{D_i}}
 \end{array}
 \label{eq:soft-deltad-table}
\end{equation}}
For the first two generations ($a=1,2$) the last column is negligible and is
dropped, exactly as in ref.~\cite{damsbsoft}. 

Using $g_Y^2=(3/5)g_1^2$ and $G_Y=(5/3)G_1$, the hypercharge weights convert
as $w_Y^{(i)}G_Y\alpha_Y^2=(3/5)w_Y^{(i)}G_1\alpha_1^2$. That is,  for $(Q,u^c,d^c,L,e^c,N^c,H_u,H_d,S,D,D^c)$,
\begin{equation}
 \frac35w_Y=\Bigl(\frac1{30},\frac8{15},\frac2{15},\frac3{10},\frac65,0,
                 \frac3{10},\frac3{10},0,\frac2{15},\frac2{15}\Bigr).
 \label{eq:soft-wY-gut}
\end{equation}

Without messenger-matter interactions, $\delta_I^{\rm Yukawa}=\delta_G^{\rm Yukawa}=0$.
On the other hand, $\delta_G^{\rm gauge}\neq0$ because the messengers
carry $SU(3)_c\tm SU(2)_L\tm U(1)_Y\tm U(1)_X$ quantum number. We have
\begin{equation}
 \delta_I^{\,i}=\frac{2d\,F_\phi^2}{(16\pi^2)^2}\,{\cal J}_i,\qquad
 \delta_G^{\,i}=\frac{d^2F_\phi^2}{(16\pi^2)^2}\,{\cal J}_i,\qquad
 \delta_G^{\,i}=\frac{d}{2}\,\delta_I^{\,i}.
 \label{eq:soft-dIG-gauge}
\end{equation}
with
\begin{equation}
 {\cal J}_i\equiv 2C_3(R_i)\Delta b_3\,g_3^4+2C_2(R_i)\Delta b_2\,g_2^4
   +2\Bigl[\Delta B_{YY}u_{i,Y}^2+2\Delta B_{YX}u_{i,Y}u_{i,X}
          +\Delta B_{XX}u_{i,X}^2\Bigr],
 \label{eq:soft-Jdef}
\end{equation}
Inserting $\Delta b_2=\Delta b_3=N$ and $\Delta B={\rm diag}(5N/3,\Delta b_X)$ with
$\Delta B_{YX}=0$,
\begin{equation}
 {\cal J}_i=2N\Bigl[C_3(R_i)g_3^4+C_2(R_i)g_2^4
            +\frac53u_{i,Y}^2+40\,u_{i,X}^2\Bigr].
 \label{eq:soft-J-expl}
\end{equation}

In the diagonal limit
\begin{equation}
 {\cal J}_i\Big|_{g_{YX}=0}
 =N\Bigl[w_3^{(i)}g_3^4+w_2^{(i)}g_2^4+\frac53w_Y^{(i)}g_Y^4
         +40\,w_X^{(i)}g_X^4\Bigr],
 \label{eq:soft-J-diag}
\end{equation}
So, the coefficients for each superfield are given by
{\small
\begin{equation}
 \begin{array}{c|rrrr|rr|l}
  & w_3 & w_2 & w_Y & w_X & \frac53w_Y & 40w_X
  & {\cal J}_i/N\;(g_{YX}=0)\\ \hline
 \widetilde Q_a    & \frac83 & \frac32 & \frac1{18} & 2
   & \frac5{54} & 80
   & \frac83g_3^4+\frac32g_2^4+\frac5{54}g_Y^4+80g_X^4\\
 \widetilde u^c_a  & \frac83 & 0 & \frac89 & 2
   & \frac{40}{27} & 80
   & \frac83g_3^4+\frac{40}{27}g_Y^4+80g_X^4\\
 \widetilde d^c_a  & \frac83 & 0 & \frac29 & 2
   & \frac{10}{27} & 80
   & \frac83g_3^4+\frac{10}{27}g_Y^4+80g_X^4\\
 \widetilde L_a    & 0 & \frac32 & \frac12 & 2
   & \frac56 & 80
   & \frac32g_2^4+\frac56g_Y^4+80g_X^4\\
 \widetilde e^c_a  & 0 & 0 & 2 & 2
   & \frac{10}3 & 80
   & \frac{10}3g_Y^4+80g_X^4\\
 \widetilde N^c_a  & 0 & 0 & 0 & 2
   & 0 & 80
   & 80\,g_X^4\\
 \widetilde H_u    & 0 & \frac32 & \frac12 & 8
   & \frac56 & 320
   & \frac32g_2^4+\frac56g_Y^4+320g_X^4\\
 \widetilde H_d    & 0 & \frac32 & \frac12 & 8
   & \frac56 & 320
   & \frac32g_2^4+\frac56g_Y^4+320g_X^4\\
 \widetilde S      & 0 & 0 & 0 & 32
   & 0 & 1280
   & 1280\,g_X^4\\
 \widetilde D_i    & \frac83 & 0 & \frac29 & 8
   & \frac{10}{27} & 320
   & \frac83g_3^4+\frac{10}{27}g_Y^4+320g_X^4\\
 \widetilde D^c_i  & \frac83 & 0 & \frac29 & 8
   & \frac{10}{27} & 320
   & \frac83g_3^4+\frac{10}{27}g_Y^4+320g_X^4
 \end{array}
 \label{eq:soft-Jtable}
\end{equation}}
The largest new effect
 is on the
 $X$-charged singlet, ${\cal J}_{\widetilde S}/N=1280\,g_X^4$:
\begin{equation}
 \delta_G^{\widetilde S}=\frac{1280\,Nd^2F_\phi^2}{(16\pi^2)^2}\,g_X^4
 =2q_S^2\,\Delta b_X\Bigl(\frac{\alpha_X}{4\pi}\Bigr)^{\!2}(dF_\phi)^2,
 \qquad \delta_I^{\widetilde S}=\frac2d\,\delta_G^{\widetilde S}.
 \label{eq:soft-deltaG-S}
\end{equation}
This has no counterpart in the NMSSM, which is the dominant correction to
$m^2_{\widetilde S}$ at large $d$, and feeds directly into $v_s$ and hence
into $\mu_{\rm eff}$ of eq.~\eqref{eq:mueff}.

Knowing the expression of soft SUSY breaking parameters at the messenger scale,
We need to RGE evolve their values from the messenger scale to the $U(1)_X$ breaking scale first, and then evolve to the SUSY scale, with the one-loop renormalization group equations obtained with SARAH~4.15.5.

For an Abelian coefficient $b_X$
\begin{equation}
 \frac{1}{g_X^2(\Lambda)}=
 \frac{1}{g_X^2(Q_0)}-
 \frac{b_X}{8\pi^2}\ln\frac{\Lambda}{Q_0}.
 \label{eq:x-running}
\end{equation}
 The value $b_X=128$ follows from the field content of
 eq.~\eqref{eq:wlow1} with the charges \eqref{eq:gut-uniform-charges}
 normalized to $q=1$, where $q_X(S)=4$ dominates the sum. Requiring $\Lambda\gtrsim10^{16}\GeV$ gives
 $g_X\lesssim0.13$, i.e.\ $g_Xq_S\lesssim0.5$, where the running
 coefficient is $b_X=128$ below $M_{\rm mess}$ and $128+40N$ above it.

\begin{table}[tbp]
\centering\small
\begin{tabular}{|l|r|r|r|r|}
\toprule
Regime &  $m_{3/2}$ [GeV] & $\Gamma_{3/2}$ [GeV] & $\tau$ [s] & $T_d$\\\hline\midrule
I & $1.62\!\times\!10^5$ & $8.5\!\times\!10^{-23}$ & $7.7\!\times\!10^{-3}$ & $13.8\MeV$\\\hline
II & $3.58\!\times\!10^5$ & $9.2\!\times\!10^{-22}$ & $7.1\!\times\!10^{-4}$ & $45.6\MeV$\\\hline
III & $3.23\!\times\!10^7$ & $6.8\!\times\!10^{-16}$ & $9.7\!\times\!10^{-10}$ & $39.1\GeV$\\
\bottomrule
\end{tabular}
\caption{Gravitino-decay estimates for the reference soft-spectrum
inputs. Decay-channel multiplicities and reheating conventions require
matching before application to another spectrum.}
\label{tab:gravitino-ref}
\end{table}

\section{Entropy relations and cosmological benchmark inputs}
\label{app:dilution}

\subsection{Thermal evolution and example source inputs}

With $x=m_\chi/T$, the dark matter comoving number density can be obtained via
equations
\begin{align}
 \frac{\mathrm dY}{\mathrm dx}
 &=-\frac{s\langle\sigma v\rangle_{\rm eff}}{Hx}
       (Y^2-Y_{\rm eq}^2),\nonumber\\
 H^2&=\frac{\rho_{\rm rad}+m_XY_X^{\rm before}s+m_\chi Ys}
             {3M_{\rm Pl}^2}.
 \label{eq:reference-boltzmann}
\end{align}
 Equation~\eqref{eq:reference-boltzmann}
accounts for the source energy in the expansion rate, but it has no
explicit decay-injection term. Applying eq.~\eqref{eq:yield} afterwards
must not be described as a full continuous-decay and re-annihilation
solution. 

The illustrative $500\GeV$ source point uses
$m_X=10^6\GeV$, $Y_X^{\rm before}=10^{-9}$ and
$B_{\rm LSP}=10^{-3}$, with reported $\Delta=1.26$ and
$\Omega_\chi h^2=0.129$. Near $1.1\TeV$, the reference choice
$Y_X^{\rm before}=1.1\times10^{-8}$ and $B_{\rm LSP}=10^{-4}$
gives reported $\Delta=4.28$ and $\Omega_\chi h^2=0.105$.
These are near-target benchmark outputs, not exact the
$\Omega_\chi h^2=0.12$ benchmark solutions.

For the heavy regime, the reference inputs are
\begin{equation}
 (m_\chi,m_X,Y_X^{\rm before},T_d)=
 (10^5\GeV,10^{10}\GeV,8.28\times10^{-9},4\MeV),
\end{equation}
The corresponding source-domination temperature is approximately $110\GeV$,
well below the reference freeze-out scale $T_f\simeq5\TeV$. The
source-to-radiation ratio at freeze-out is $2.2\times10^{-2}$.
The produced dilution $2.8\times10^{4}$ and the requirement
$(Y_{\rm th}+B_{\rm LSP}Y_X^{\rm before}+Y_{3/2}^{\rm th})/Y_{\rm req}=2.8\times10^{4}$ agree at the precision relevant here.

The source abundance, long lifetime and LSP multiplicity are separate
requirements, not consequences of the dAMSB mass formula. The abundance
is not free once the dilution is fixed: the entropy budget of
eq.~\eqref{eq:delta-def} requires
$Y_X^{\rm before}=u/m_X$ with $u\simeq\frac34\Delta_d T_d\simeq82.8\,\GeV$
for the heavy benchmark, as recorded in
eq.~\eqref{eq:required-YX}.

\subsection{An alternative scalar entropy source}
\label{sec:sigma}

The main text separates two requirements on a late entropy source: the
entropy budget, which fixes $\Delta_d$ and hence $Y_X^{\rm before}$, and
the LSP multiplicity $B_{\rm LSP}$, which decides whether the same decay
refills the abundance that it dilutes. A source that meets the second
requirement by construction is a gauge-singlet, $R$-even scalar $\sigma$
with gravitational-strength visible decays. If the Higgsino is the LSP
and
\begin{equation}
 m_\sigma<2m_\chi,
\end{equation}
production of a pair of on-shell superpartners is kinematically forbidden.
This removes direct LSP production by the source within the assumed
particle spectrum, so that $B_{\rm LSP}=0$ follows from the spectrum
rather than from an operator assignment. It does not follow solely from
deflected anomaly mediation; the scalar and its initial displacement are
additional cosmological ingredients. Related heavy-Higgsino cosmologies
motivate this possibility~\cite{Langhoff}.

The decay width of $\sigma$ is
\begin{equation}
 \Gamma_\sigma=c\frac{m_\sigma^3}{M_{\rm Pl}^2},\qquad
 T_d=1.74\,g_*^{-1/4}\sqrt{\Gamma_\sigma M_{\rm Pl}}.
 \label{eq:sigma-decay}
\end{equation}
With the reduced Planck mass $M_{\rm Pl}=2.435\times10^{18}\GeV$, $c=1$ and the reference $g_*=10.75$ give the mass interval
approximately $35\TeV\lesssim m_\sigma<200\TeV$, corresponding to
$4\MeV\lesssim T_d<55\MeV$. The upper bound is the pair-production
threshold for $m_\chi=100\TeV$. The initial displacement $\varphi_0$ is the field value of
$\sigma$ at the onset of coherent oscillations ($H\simeq m_\sigma$). The
stored energy density $\rho_\sigma\simeq\tfrac12 m_\sigma^2\varphi_0^2$
redshifts as matter until the decay at $T_d$, and its value is fixed by the
required dilution $\Delta_d$.  The value required for the
benchmark dilution is of order $10^{12}\GeV$ for $\Delta_d\simeq6\times10^{3}$
and scales as $\sqrt{\Delta_d}$, i.e.\ as $3\times10^{12}\,\GeV$ for
the benchmark of eq.~\eqref{eq:heavy-benchmark}.

Table~\ref{tab:sigma} shows the correlated values of the gravitational
source parameters. Its $200\TeV$ entries show the pair-production
threshold, where a kinematically blocked benchmark must lie strictly below
that value for $m_\chi=100\TeV$. The approximation of a common $g_*=10.75$ and
instantaneous reheating is used for the table.

\begin{table}[tbp]
\centering\small
\begin{tabular}{|r|r|r|r|}
\toprule
$m_\sigma$ [TeV] & $T_d$ [MeV] & $c$ & $\varphi_0$ [GeV]\\
\midrule
$34.81$ & $4.03$ & $1$ & $2.26\times10^{12}$\\
$50$ &  $6.89$ & $1$ & $2.711\times10^{12}$\\
$100$ &$4$ & $0.0422$ & $1.74\times10^{12}$\\
$200$ (threshold) & $4$ & $0.00527$ & $1.46\times10^{12}$\\
\bottomrule
\end{tabular}
\caption{Reference scalar-source inputs for the heavy Higgsino regime.
The first block fixes $c=1$; the second fixes $T_d=4\MeV$ and determines
$c=(T_d/T_d^{c=1})^2$.}
\label{tab:sigma}
\end{table}

\section{Experimental inputs and exploratory solar estimates}
\label{app:phenodata}

Table~\ref{tab:dd-inputs} records the experimental inputs cited in the
reference analysis. The observed upper limit and the median expected
sensitivity are different quantities. No new data extraction or
likelihood reconstruction is implied by this table.

\begin{table}[tbp]
\centering\small
\begin{tabularx}{\linewidth}{Yrrr}
\toprule
Data set & Exposure [t\,yr] & Best mass [GeV] & Observed SI limit [cm$^2$]\\
\midrule
LZ first result~\cite{LZ2023} & $0.9$ & $36$ & $9.2\times10^{-48}$\\
LZ later result~\cite{LZ42} & $4.2\pm0.1$ & $40$ & $2.2\times10^{-48}$\\
XENONnT first result~\cite{Xenonnt2023} & $\simeq1.1$ & $28$ & $2.58\times10^{-47}$\\
XENONnT later result~\cite{Xenonnt31} & $3.1$ & $30$ & $1.7\times10^{-47}$\\
\bottomrule
\end{tabularx}
\caption{Quoted 90\% confidence-level spin-independent WIMP--nucleon
limits at each result's best mass. These are not the limits at the
three Higgsino benchmark masses.}
\label{tab:dd-inputs}
\end{table}

The $4.2\,\mathrm{t\,yr}$ LZ exposure corresponds to $280$ live days
in the cited result. Its quoted median sensitivity is
$5.1\times10^{-48}\,\mathrm{cm}^2$. The $3.1\,\mathrm{t\,yr}$
XENONnT result quotes a median sensitivity
$1.4\times10^{-47}\,\mathrm{cm}^2$ at $41\GeV$.
Mass-dependent data at $500\GeV$, $1.1\TeV$ and $100\TeV$ must be
extracted from the corresponding releases before forming a predicted-to-
excluded cross-section ratio. 
\subsection{Solar transport and amplitude suppression}

For the 1.1 TeV Higgsino dark matter interpretation, the reference solar studies
quote the limits near $506\keV$, extending to approximately $566\keV$
with additional scattering effects, and an independent limit near
$557\keV$~\cite{Pospelov,Bose}. These limits depend on how each analysis treats capture,
thermalization and annihilation, and should be read together with those
assumptions. Under these equilibrium assumptions, no part of the
low-splitting event window survives.

Some transport calculation in the references quotes
a thermalization scale $\sigma_{\rm SI}\simeq4.96\times10^{-46} 
\,\mathrm{cm}^2$  (which is the critical cross section at which the behaviour of
the trapped population changes from promptly thermalized to ballistic) with a broad estimated interval $3\times10^{-47}$--$1.6\times10^{-44}\,\mathrm{cm}^2$. 
 In the capture-annihilation equilibrium treatment for solar dark matter, the predicted signal exceeds the bound by a factor $\sim2.2\times10^{4}$. Escaping it would require
suppressing the elastic amplitude by a factor $\sim0.0067$.
 These are estimations, not a published solar
likelihood or a demonstrated blind-spot (cancellation) point in the
neutralino sector. A full orbit-averaged transport calculation may still be needed to eliminate the large uncertainty.

The $100\TeV$ Higgsino scenario reduces the number-density of dark matter and uses a quoted equilibration time $1.2\times10^9$ years (compared with a solar age of $4.6\times10^9$ years. Fixing a mass-dependent exclusion boundary requires the
capture rate, the annihilation volume, the branching fractions and the
neutrino propagation. Tentative splitting estimates in the $0.40$--$0.46\MeV$ range by reading the figures are therefore not promoted to limits here. The high-energy sideband and the halo and nuclear systematics likewise require their own likelihood inputs
before a joint survival contour can be drawn~\cite{Rodd,Baxter}.

\end{document}